\documentclass[AMA,STIX1COL]{WileyNJD-v2}

\articletype{Research Article}%

\received{}
\revised{}
\accepted{}
        
\begin{document}

\title{Analysis and optimal individual pitch control decoupling by inclusion of an azimuth offset in the multi-blade coordinate transformation}

\author[1]{Sebastiaan Paul Mulders}
\author[1]{Atindriyo Kusumo Pamososuryo}
\author[1]{Gianmarco Emilio Disario}
\author[1]{Jan-Willem van Wingerden}

\authormark{S.P. MULDERS \textsc{et al}}

\address[1]{\orgdiv{Delft Center for Systems and Control, Faculty of Mechanical Engineering}, \orgname{Delft University of Technology}, \orgaddress{\country{The Netherlands}}}

\corres{Sebastiaan Paul Mulders, Delft Center for Systems and Control, Faculty of Mechanical Engineering, Mekelweg 2, 2628 CD Delft, The Netherlands.\\ \email{s.p.mulders@tudelft.nl}}


\abstract[Abstract]{With the trend of increasing wind turbine rotor diameters, the mitigation of blade fatigue loadings is of special interest to extend the turbine lifetime. Fatigue load reductions can be partly accomplished using Individual Pitch Control (IPC) facilitated by the so-called Multi-Blade Coordinate (MBC) transformation. This operation transforms and decouples the blade load signals in a yaw- and tilt-axis. However, in practical scenarios, the resulting transformed system still shows coupling between the axes, posing a need for more advanced Multiple-Input Multiple-Output (MIMO) control architectures. This paper presents a novel analysis and design framework for decoupling of the non-rotating axes by the inclusion of an azimuth offset in the reverse MBC transformation, enabling the application of simple Single-Input Single-Output (SISO) controllers. A thorough analysis is given by including the azimuth offset in a frequency-domain representation. The result is evaluated on simplified blade models, as well as linearizations obtained from the NREL~5\nobreakdash-MW reference wind turbine. A sensitivity and decoupling assessment justify the application of decentralized SISO control loops for IPC. Furthermore, closed-loop high-fidelity simulations show beneficial effects on pitch actuation and blade fatigue load reductions.}


\maketitle
\section{Introduction}
As wind turbine blades are getting larger and more flexible with increased power ratings, the need for fatigue load reductions is getting ever stronger \citep{ref:caselitz1997reduction}. For a large Horizontal Axis Wind Turbine (HAWT), the wind varies spatially and temporally over the rotor surface due to the combined effect of turbulence, wind shear, yaw-misalignment and tower shadow \citep{Upwind2006Report}, and give rise to periodic blade loads. The blades itself mainly experience a once-per-revolution $1$P cyclic load, whereas the tower primarily experiences a $3$P cyclic load in the case of a three-bladed wind turbine.

To reduce fatigue loadings, the capability of wind turbines to individually pitch its blades is exploited by Individual Pitch Control (IPC). The pitch contributions for fatigue load reductions are generally formed with use of the azimuth-dependent Multi-Blade Coordinate (MBC) transformation, acting on out-of-plane blade load measurements. The forward MBC transformation transforms the load signals from a rotating into a non-rotating reference frame, resulting in tilt and yaw rotor moments. After the obtained signals have been subject to control actions, the reverse MBC transformation is used to obtain implementable individual pitch contributions. The MBC transformation is also used in other fields such as in electrical engineering where it is often referred to as the Park or direct-quadrature-zero (dq0) transformation \citep{park1929two}, and in helicopter theory where it is called the Coleman transformation \citep{johnson2012helicopter}.

IPC for wind turbine blade fatigue load reductions using the MBC transformation is widely discussed in the literature \citep{ref:Menezes2018WTCReview}. While high-fidelity simulation software shows promising results and field tests have been performed \citep{ref:bossanyi2013validation, ref:solingen2016field}, the in-field deployment of IPC is still scarce, likely due to the increased pitch actuator loading by continuous operation of IPC \citep{ref:shan2013field}. Also, due to the complicated maintenance of blade load sensors, research has been conducted on load estimation using measurements from the turbine fixed tower support structure \citep{ref:jelavic2010estimation}. In research, various IPC control methodologies have been proposed such as a comparison of more advanced Linear-Quadratic-Gaussian (LQG) and simple Proportional-Integral (PI) control \citep{ref:bossanyi2003individual}, application of $H_\infty$ techniques \citep{ref:geyler2007individual}, Repetitive Control (RC) \citep{ref:navalkar2014sprc} and Model Predictive Control (MPC) using short-term wind field predictions \citep{ref:spencer2013model}. The effect of pitch errors and rotor asymmetries and imbalances is also investigated \citep{ref:petrovic2015advanced}. 

Common in industry is to apply an azimuth offset in the reverse MBC transformation, however, its interpretation, analysis and effect is more than ambiguous. Bossanyi \citep{ref:bossanyi2003individual} states that a constant offset can be added to account for the remaining interaction between the two transformed axes. Later, the same author suggests \citep{bossanyi2009upwind} that a small offset in the reverse transformation can be used to account for the phase lag between the controller and pitch actuator. Houtzager et al. \citep{houtzager2013wind} states that the performance of IPC is reduced by a large phase delay between the controller and pitch actuator, but that also the total phase lag of the open-loop system at the $1$P and $2$P harmonics can be compensated for by including the offset. Mulders \citep{mulders2015iterative} shows that the azimuth offset changes the dynamics of the IPC signal and that an optimum is present in terms of Damage Equivalent Load (DEL). During field tests on the three-bladed Control Advanced Research Turbines (CART3) \citep{ref:bossanyi2013validation}, it is noted that for successful attenuation of the $1$P and $2$P harmonics, distinct offsets are needed for both frequencies: the offset values are found experimentally and are said to possibly reflect the frequency dependency of the pitch actuator. The same paper also reveals that the azimuth offset is required to compensate for cross-coupling between the fixed-frame axes. The work of Solingen et al. \citep{ref:solingen2016field} mentions that the MBC transformation can incorporate compensation for phase delays by including an azimuth offset in the reverse transformation.

All of the papers discussed above impose different claims on the effect of the azimuth offset in the reverse transformation, but in none of these papers a thorough analysis is given. Coupling between the tilt and yaw axes is demonstrated \citep{lu2015analysis} by a frequency-domain analysis of the MBC transformation with simplified control-oriented blade models. It is stated that this coupling should be taken into account during controller design and a $\mathcal{H}_\infty$ loop-shaping approach is therefore employed. However, the authors do not consider the effect of the azimuth offset in their derivation for decoupling of the non-rotating axes, and the resulting possible implementation of IPC with SISO controllers. The cross-coupling of the transformed system is taken into account in Ungur\'{a}n et al. \cite{ref:Unguran2019FFIPC} by matrix-multiplication with the steady-state gain of the inverse plant. Doing so enables the application of an IPC controller with decoupled SISO control loops, however, requires evaluation of the low-frequent diagonal and off-diagonal frequency responses. The latter might be challenging from a numerical as well as a practical perspective.

This paper uses the azimuth offset for decoupling of the transformed system, and gives a thorough analysis on the effect by providing the following contributions:
\begin{itemize}
	\setlength{\itemsep}{0pt}
	\setlength{\parskip}{0pt}
	\setlength{\parsep}{0pt}
	\item Providing a formal frequency-domain framework for analysis of the azimuth offset;
	\item Describing a design methodology to find the optimal offset angles throughout the entire turbine operating envelope;
	\item Demonstrating the approach for rotor models of various fidelity, and thereby showing the implications on the accuracy of the found optimal offset;
	\item Showcasing the effects of the azimuth offset using simplified blade models;
	\item Performing an assessment on the degree of decoupling using the Gershgorin circle theorem and the consequences for controller synthesis by analysis of the sensitivity function;
	\item Using closed-loop high-fidelity simulations to show the offset implications on pitch actuation and blade load signals.
\end{itemize}

This paper is organized as follows. In Section~\ref{sec:MBC}, the time-domain MBC representation incorporating the azimuth offset is presented, and is used in an open-loop setting to formalize the problem by an illustrative example using the NREL~5\nobreakdash-MW reference wind turbine. Next, in Section~\ref{sec:MBC_FreqDomain}, a frequency-domain representation of the MBC transformation including the offset is derived. Two distinct rotor model structures are proposed, including and excluding blade dynamic coupling. The two beforementioned model structures are employed in Section~\ref{sec:LA} to show the effect of the offset on simplified blade models. Subsequently, in Section~\ref{sec:NLA}, the results are evaluated on linearizations of the NREL~5\nobreakdash-MW turbine and validated to results presented in the first section. In Section~\ref{sec:ASS}, an assessment on a control design with diagonal integrators and the effectiveness in terms of decoupling is given. In Section~\ref{sec:SIMU}, closed-loop high-fidelity simulations are performed to show the implications on pitch actuation and blade fatigue loading. Finally, conclusions are drawn in Section~\ref{sec:CONCLUSIONS}.

\section{Time domain Multi-Blade Coordinate transformation and problem formalization}\label{sec:MBC}
This section starts with the time-domain formulation of the MBC transformation, including the option for an azimuth offset in the reverse transformation. Next, Section~\ref{sec:MBC_ProblemStatement} shows high-fidelity simulation results of the NREL~5\nobreakdash-MW turbine to showcase the effect of the offset. The results formalize the problem and are a basis for further analysis in subsequent sections.
\begin{figure}[b!]
	\centering
	\includegraphics[scale=1.0]{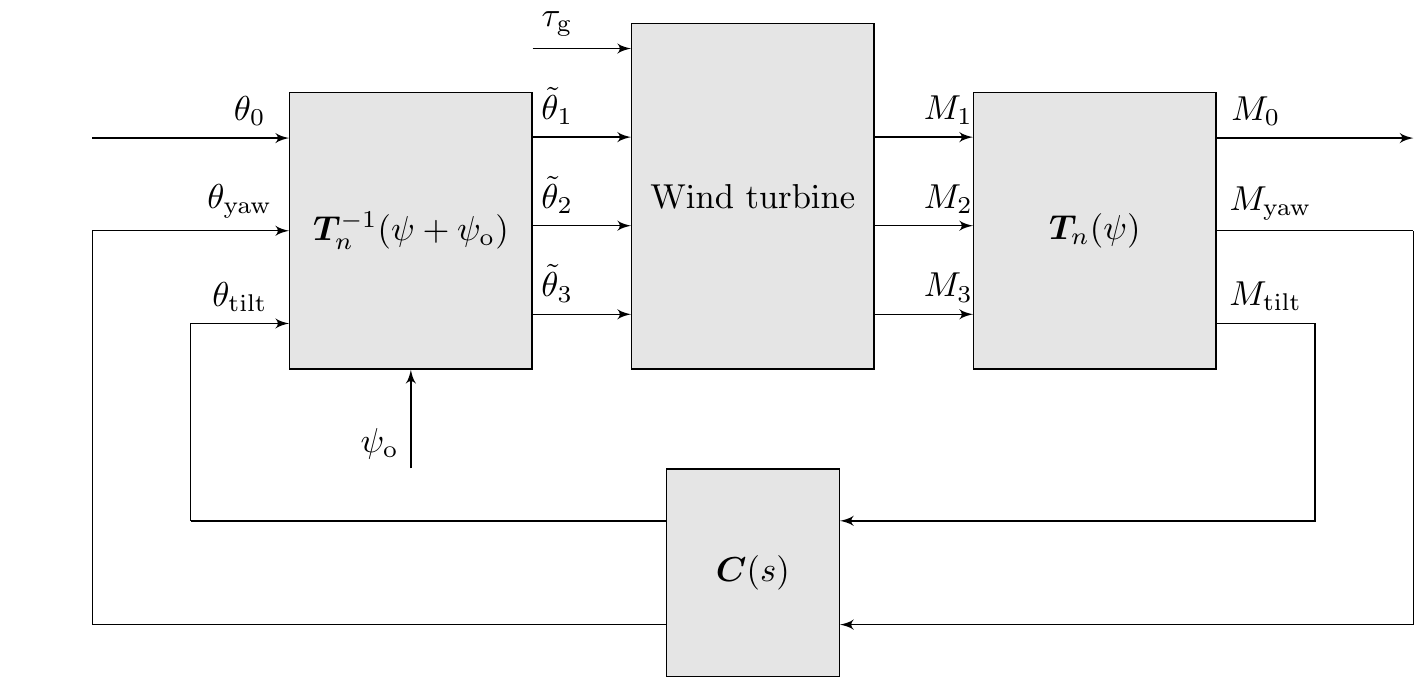}
	\caption{Typical implementation of IPC using azimuth-dependent forward and reverse MBC transformations $\boldsymbol{T}(\psi)$ and $\boldsymbol{T}^{-1}(\psi+\psi_\mathrm{o})$, decoupling and transforming blade load harmonics to a fixed reference frame. The IPC controller $\boldsymbol{C}(s)$ generates the fixed-frame pitch contributions by acting on the tilt and yaw moments. The non-rotating signals are transformed back to the rotating frame by the reverse transformation, resulting in pitch contributions $\tilde{\theta}_b$, made up of collective and individual pitch contributions $\theta_0$ and $\theta_b$, respectively. The generator torque control signal is indicated by ${\tau}_\mathrm{g}$. The collective pitch and generator torque control signals are generated by turbine controllers, omitted in this figure.}%
	\label{fig:MBC_ClosedLoop}%
\end{figure}

\subsection{Time domain MBC representation}\label{sec:MBC_General}
Conventional implementations of IPC use the MBC transformation for fatigue load reductions. The MBC transformation transforms measured blade moments from a rotating reference frame to a non-rotating frame, and decouples the signals for convenient analysis and controller design. A schematic diagram of the general IPC configuration for a three-bladed wind turbine is presented in Figure~\ref{fig:MBC_ClosedLoop}, where the generator torque and collective pitch angle control signals are indicated by $\tau_\mathrm{g}$ and $\theta_0$, respectively. The relations transforming the rotating out-of-plane blade moments $M_b$, to their respective non-rotating degrees of freedom \citep{ref:bir2008MBC} are defined by the forward MBC transformation
\begin{align}
\left[ \begin{array}{c} M_0(t) \\ M_\mathrm{tilt}(t) \\ M_\mathrm{yaw}(t) \end{array} \right] &= \underbrace{\frac{2}{B}\left[ \begin{array}{c c c} 1 & 1 & 1 \\ \cos{\left(n\psi_1(t)\right)} & \cos{\left(n\psi_2(t)\right)} & \cos{\left(n\psi_3(t)\right)} \\ \sin{\left(n\psi_1(t)\right)} & \sin{\left(n\psi_2(t)\right)} & \sin{\left(n\psi_3(t)\right)} \\ \end{array} \right]}_{\boldsymbol{T}_{n}(\psi(t))} \left[ \begin{array}{c} M_{1}(t) \\ M_{2}(t) \\ M_{3}(t) \end{array} \right] \label{eqn:MBCtransform-2B-matforw},
\end{align}
where $n\in\mathbb{Z}^+$ is the harmonic number, $B=3$ the total number of blades, and $\psi_b\subset\mathbb{R}$ is the azimuth position of blade $b\subset\mathbb{Z}^+$ with respect to the reference azimuth $\psi$, given by
\begin{align}
\psi_b(t) &= \psi(t) + (b-1)\frac{2\pi}{B},
\label{eq:1azimuth_b}
\end{align}
and the rotor azimuth coordinate system is defined as $\psi_b = 0$ when the blade is in the upright vertical position.

The obtained non-rotating (fixed-frame) degrees of freedom are called rotor coordinates because they represent the cumulative behavior of all rotor blades. The collective mode $M_0$ represents the combined out-of-plane flapping moment of all blades. The cyclic modes $M_\mathrm{tilt}$ and $M_\mathrm{yaw}$ respectively represent the rotor fore-aft tilt (rotation around a horizontal axis and normal to the rotor shaft) and the rotor side-side coning (rotation around a vertical axis and normal to the rotor shaft) \citep{ref:bir2008MBC}. The cyclic modes are most important because of their fundamental role in the coupled motion of the rotor in the non-rotating system. For axial wind flows the collective and cyclic modes of the rotor degrees of freedom couple with the fixed system. 

After control action by the IPC controller $\boldsymbol{C}(s)\equiv\left\{C_{ij}(s)\right\}_{2\times2}$, the reverse transformation converts the obtained non-rotating pitch angles $\theta_\mathrm{tilt}$ and $\theta_\mathrm{yaw}$ in the non-rotating frame back to the rotating frame
\begin{align}
\left[ \begin{array}{c} \tilde{\theta}_{1}(t) \\ \tilde{\theta}_{2}(t) \\ \tilde{\theta}_{3}(t) \end{array} \right] &= \left[\begin{array}{c} {\theta}_{0} + \theta_1 \\ {\theta}_{0} + \theta_2 \\ {\theta}_{0} + \theta_3 \end{array}\right] =  \underbrace{\left[ \begin{array}{ccc} 1 & \cos{\left[n\left(\psi_1(t)+\psi_\mathrm{o}\right)\right]} & \sin{\left[n\left(\psi_1(t)+\psi_\mathrm{o}\right)\right]} \\ 1 & \cos{\left[n\left(\psi_2(t)+\psi_\mathrm{o}\right)\right]} & \sin{\left[n\left(\psi_2(t)+\psi_\mathrm{o}\right)\right]} \\ 1 & \cos{\left[n\left(\psi_3(t)+\psi_\mathrm{o}\right)\right]} & \sin{\left[n\left(\psi_3(t)+\psi_\mathrm{o}\right)\right]} \end{array} \right]}_{\boldsymbol{T}_{n}^{-1}(\psi(t)+\psi_\mathrm{o})}	\left[ \begin{array}{c} \theta_0(t) \\ \theta_\mathrm{tilt}(t) \\ \theta_\mathrm{yaw}(t) \end{array} \right],	
\label{eq:1MBCTransformInv}
\end{align}
where the resulting pitch angle $\tilde{\theta}_i$ consists of collective pitch and IPC contributions $\theta_0$ and $\theta_i$, respectively, and the azimuth offset is represented by $\psi_\mathrm{o}\in\mathbb{R}$. The offset could have also been incorporated in the forward transformation and an extensive analysis on this aspect is given in Disario \citep{ref:Disario2018MScThesis}.

The main topic of this paper is to perform a thorough analysis on the effects of the offset and to provide a framework for derivation of the optimal phase offset throughout the complete turbine operating envelope. The analysis is performed on the $1$P rotational frequency, however, the framework given is applicable to all $n$P harmonics.

\subsection{Problem formalization by an illustrative example}\label{sec:MBC_ProblemStatement}
\begin{figure}[!b]
	\centering
	\includegraphics[scale=1.0]{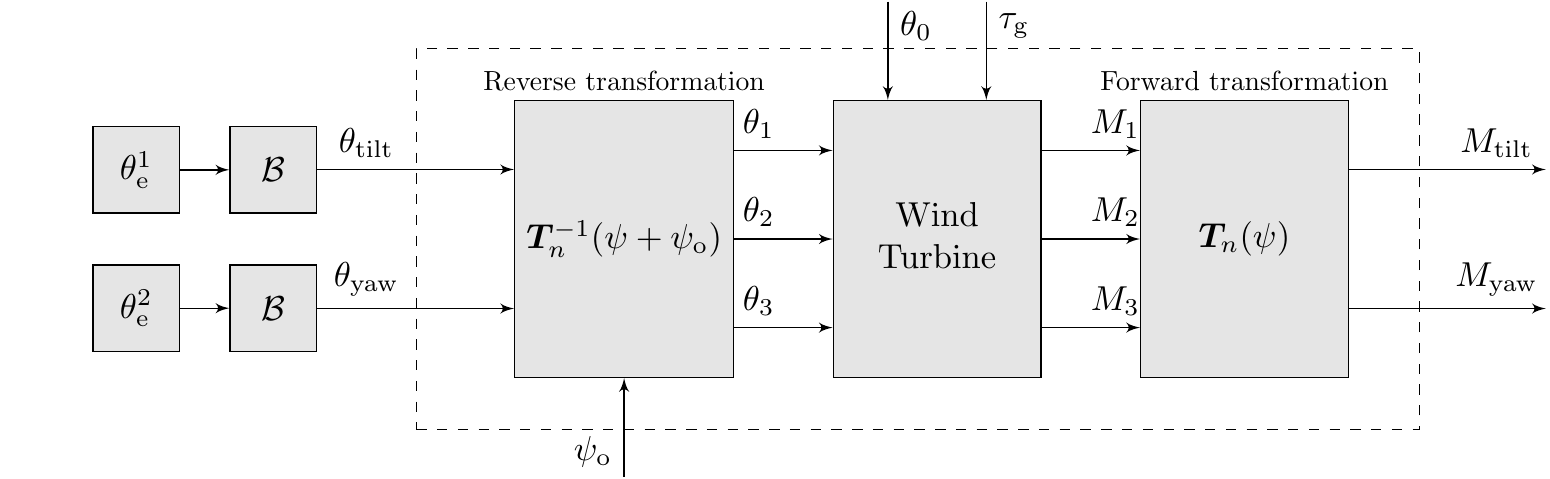}
	\caption{Set-up for identification of a non-parametric spectral model $\boldsymbol{P}_\mathrm{s}(j\omega)$ of the dashed system. The wind turbine is a non-linear model and is subject to a steady-state collective pitch angle $\theta_0$ and generator torque $\tau_\mathrm{g}$. The non-rotating pitch excitation signals $\theta_\mathrm{e}$ are filtered by a band-pass filters $\mathcal{B}$, and the wind turbine includes a pitch actuator model. The identification is performed for distinct azimuth offsets $\psi_\mathrm{o}$.}%
	\label{fig:MBC_OpenLoop_Spectral}%
\end{figure}

To showcase the effect of the azimuth offset, the implementation depicted in Figure~\ref{fig:MBC_OpenLoop_Spectral} is used to identify non-parametric spectral models of the system indicated by the dashed box for different offsets and wind speeds. To this end, the NREL~5\nobreakdash-MW reference turbine is subject to the previously introduced MBC transformation, implemented in an open-loop set-up using FAST (Fatigue, Aerodynamics, Structures, and Turbulence): a high-fidelity open-source wind turbine simulation software package \citep{ref:FASTv816}. The non-linear wind turbine is commanded with fixed collective pitch and generator torque demands, corresponding to a constant wind speed in the range $U = 5-25$~m\,s\textsuperscript{-1}. The forward and reverse transformations are employed at the $n=1$ ($1$P) harmonic, and the reverse transformation is configured with different offsets values. The wind turbine includes first-order pitch actuator dynamics with a bandwidth of $\omega_\mathrm{a} = 2.5$~rad\,s\textsuperscript{-1}, which results in an additional open-loop frequency-dependent phase loss.

For identification purposes, the excitation signals $\theta_\mathrm{e}^{i}$ are taken as Random Binary Signals (RBS) of different seeds with an amplitude of $1$~deg and a clock period \citep{ref:Ljung1999SysID} of $N_\mathrm{c} = 1$, resulting in flat signal spectra. A bandpass filter $\mathcal{B}$ is included to limit the low and high the frequency content entering the (pitch) system. The cut-in and cut-off frequencies of the bandpass filter are specified at $10^{-3}$ and $10^{2}$~rad\,s\textsuperscript{-1}, respectively, as results will be evaluated in the frequency range from $10^{-1}$ to $10^{1}$~rad\,s\textsuperscript{-1}. The sampling frequency is set to $\omega_\mathrm{s} = 125$~Hz, and the total simulation time is $2200$~s, where the first $200$~s are discarded to exclude transient effects from the data set. A frequency-domain estimate of the non-rotating system transfer function $\boldsymbol{P}_\mathrm{s}\in\mathbb{C}^{2\times2}$ is obtained from the tilt and yaw pitch to blade moment signals by spectral analysis\footnote{For spectral analysis, the \texttt{spa\_avf} routine of the Predictor-Based-Subspace-IDentification (PBSID) toolbox \citep{ref:PBSIDToolbox} is used.}.
\begin{figure}[t!]
	\centering
	\includegraphics[scale=1.0]{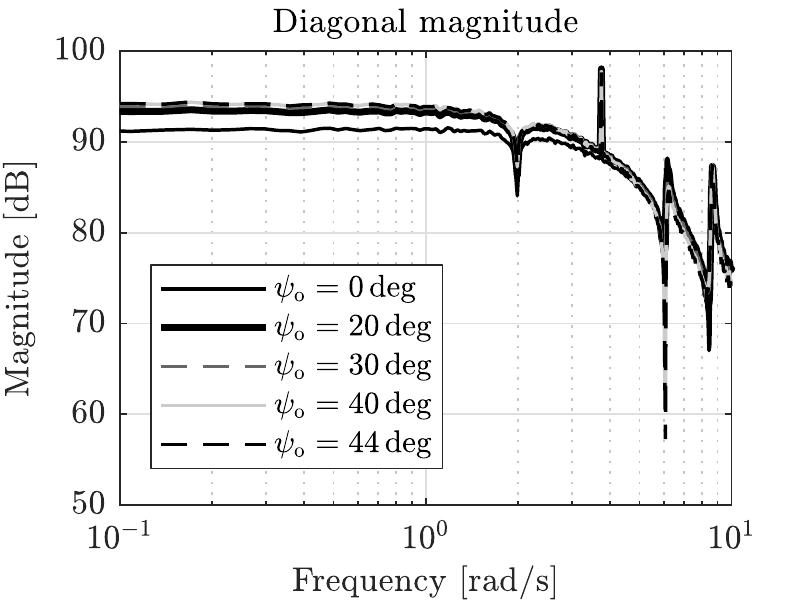}
	\includegraphics[scale=1.0]{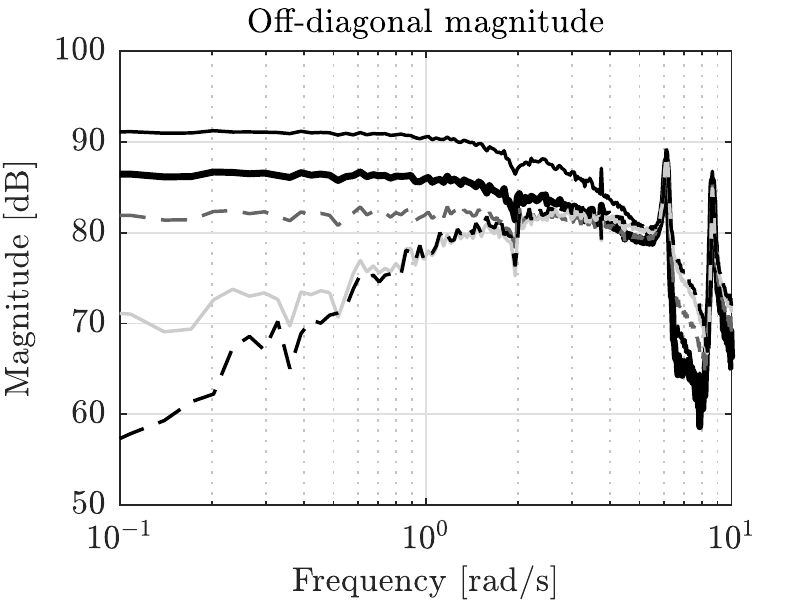}
	\caption{Diagonal and off-diagonal magnitudes of $\boldsymbol{P}_\mathrm{s}$ for the input-output pairs $\left(\theta_\mathrm{tilt},~M_\mathrm{tilt}\right)$ and $\left(\theta_\mathrm{tilt},~M_\mathrm{yaw}\right)$, obtained from non-linear wind turbine model simulations with $U=25$~m\,s\textsuperscript{-1}. The reverse MBC transformation is supplied with different azimuth offset values. It is shown that the offset primarily influences the low-frequency off-diagonal magnitude.}%
	\label{fig:MBC_ProblemStatement_BodeMag}%
\end{figure}
\begin{figure}[b!]
	\centering
	\includegraphics[scale=1.0]{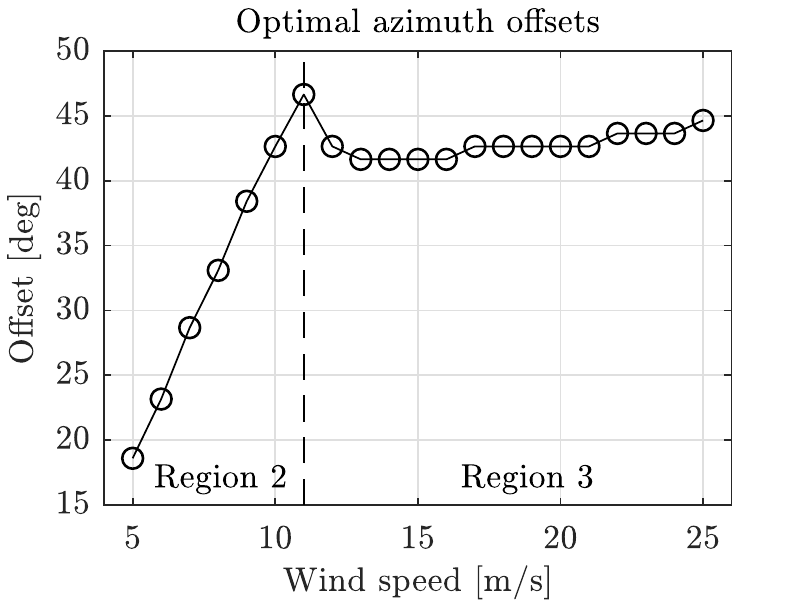}
	\caption{The optimal azimuth offset as function of wind speed, both with an accuracy up to the nearest integer value. The optimal offset minimizes $R_\mathrm{\#}$ of the frequency-domain estimate of the system transfer function. It is shown that the operating condition of the turbine has a high influence on the optimal offset value.}%
	\label{fig:MBC_ProblemStatement_OptOffset}%
\end{figure}

Figure~\ref{fig:MBC_ProblemStatement_BodeMag} presents a spectral analysis of the non-rotating system subject to a wind speed of $25$~m\,s\textsuperscript{-1} for different offset values. Because the MBC transformation moves the $1$P harmonic to a $0$P DC contribution, the aim is to minimize the off-diagonal low-frequency content. It is shown that $\psi_\mathrm{o}$ primarily influences the low-frequency magnitude from the off-diagonal terms of the 2-by-2 system. From now on, the \textit{optimal offset} is defined as the value for which the main-diagonal terms have a maximized, and off-diagonal terms have a minimized low-frequency gain. This is further formalized using the Relative Gain Array (RGA) \citep{ref:skogestad2007multivariable}, which is defined as the element-wise product (the Hadamard or Schur product, indicated by $\left(\circ\right)$) of the non-rotating system frequency response and its inverse-transpose
\begin{align}
\boldsymbol{R}(j\omega) &= \boldsymbol{P}(j\omega)\circ\left(\boldsymbol{P}(j\omega)\right)^{-\mathrm{T}}.
\end{align}
Subsequently, the level of system interaction over a frequency range is quantified by a single off-diagonal element of the RGA, defined by
\begin{align}
R_\mathrm{\#} &= \frac{1}{L}\sum_{i=1}^{L}\left| {R}_{12}(j\omega_{\mathrm{s},{i}}) \right|,
\end{align}
where ${\omega}_{\mathrm{s},i}\in\mathbb{R}$ for $i\in\left\{1,~2,~\dots,~\mathrm{L}\right\}$ specifies the frequency range of interest. In Figure~\ref{fig:MBC_ProblemStatement_OptOffset}, the optimal offset is evaluated by minimization of $R_\mathrm{\#}$ for the low-frequency range from $\omega_\mathrm{s,1}=0.1$ to $\omega_\mathrm{s,\mathrm{L}}=1$~rad\,s\textsuperscript{-1}. It is shown that the optimal offset value changes for each wind speed and is thus highly dependent on the turbine operating conditions. An elaborate analysis on the establishment of the optimal azimuth offset is given in the remainder of this paper.

\section{Frequency domain Multi-Blade Coordinate representation}\label{sec:MBC_FreqDomain}
In the work of Lu et al.\citep{lu2015analysis}, a three-bladed wind turbine incorporating the MBC forward and reverse transformations is expressed in the frequency domain using a transfer function representation. By doing so, it was found that while the assumed simplified rotor model -- consisting out of three identical linear blade models -- did not include cross-terms, coupling between the tilt- and yaw-axis was present. This chapter extends the derivation for different rotor model structures, by also including the azimuth offset.

In Sections~\ref{sec:MBC_FreqDomain_Prelim}~to~\ref{sec:MBC_FreqDomain_Reverse}, the derivation of a frequency-domain representation of the MBC transformation is presented. Sections~\ref{sec:MBC_FreqDomain_DRotMod}~and~\ref{sec:MBC_FreqDomain_CRotMod} combine the obtained results by assuming rotor model structures excluding and including cross-terms. Finally, Section~\ref{sec:MBC_AziOffset} incorporates the azimuth offset in the framework.

\subsection{Preliminaries}\label{sec:MBC_FreqDomain_Prelim}
For analysis of the considered system in the frequency domain, the rotor speed denoted by $\omega_\mathrm{r}$ is taken constant such that the azimuth is expressed as $\psi(t) = \omega_\mathrm{r}t$. The following Laplace transformations \citep{Oppenheim2013SignalsSystems} are defined first as they are used subsequently in the derivation
\begin{align}
\mathcal{L}\left\{\cos(n\omega_\mathrm{r}t)x(t)\right\} &= \mathcal{L}\left\{\frac{e^{jn\omega_\mathrm{r}t}+e^{-jn\omega_\mathrm{r}t}}{2}x(t)\right\} = \frac{1}{2}\left(X(s-jn\omega_\mathrm{r})-X(s+jn\omega_\mathrm{r})\right),\\
\mathcal{L}\left\{\sin(n\omega_\mathrm{r}t)x(t)\right\} &= \mathcal{L}\left\{\frac{\left(e^{jn\omega_\mathrm{r}t}-e^{-jn\omega_\mathrm{r}t}\right)}{2j}x(t)\right\} = \frac{1}{2j}\left(X(s-jn\omega_\mathrm{r})-X(s+jn\omega_\mathrm{r})\right),
\end{align}
where $x(t)$ is an arbitrary signal and $X(s)$ is its Laplace transform. With a slight abuse of notation, the frequency-shifted Laplace operators are defined as
\begin{align}
s_{-} &= s-jn\omega_\mathrm{r},\\
s_{+} &= s+jn\omega_\mathrm{r},
\end{align}
where $n\in\mathbb{Z}^+$ is the harmonic number and $j=\sqrt{-1}$ is the imaginary unit.

\subsection{Forward MBC transformation}\label{sec:MBC_FreqDomain_Forward}
The time-domain representation of the forward MBC transformation in Eq.~\eqref{eqn:MBCtransform-2B-matforw}, is now rewritten using trigonometric identities \citep{Stewart2009Calculus} as
\begin{align}
M_\mathrm{tilt}(t) &= \frac{2}{3}\sum_{b = 1}^{3} M_b(t)\left[\cos\left(n\omega_\mathrm{r}t\right)\cos\left(\frac{2\pi n(b-1)}{3}\right)-\sin\left(n\omega_\mathrm{r}t\right)\sin\left(\frac{2\pi n(b-1)}{3}\right)\right],\\
M_\mathrm{yaw}(t) &= \frac{2}{3}\sum_{b = 1}^{3} M_b(t)\left[\sin\left(n\omega_\mathrm{r}t\right)\cos\left(\frac{2\pi n (b-1)}{3}\right)+\cos\left(n\omega_\mathrm{r}t\right)\sin\left(\frac{2\pi n(b-1)}{3}\right)\right].
\end{align}
Now the cyclic modes are transformed to their frequency-domain representation
\begin{multline}
\begin{bmatrix}M_\mathrm{tilt}(s)\\M_\mathrm{yaw}(s)\end{bmatrix} = \frac{2}{3}
\underbrace{\frac{1}{2}
	\begin{bmatrix}
	1 & j \\
	-j & 1
	\end{bmatrix}
	\begin{bmatrix}
	\cos{(0)} & \cos{(2\pi n/3)} & \cos{(4\pi n/3)}\\
	\sin{(0)} & \sin{(2\pi n/3)} & \sin{(4\pi n/3)}
	\end{bmatrix}}_{\boldsymbol{C}_{\mathrm{L},n}}
\begin{bmatrix}
M_1(s_{-})\\
M_2(s_{-})\\
M_3(s_{-})
\end{bmatrix}\\
+
\frac{2}{3}
\underbrace{\frac{1}{2}
	\begin{bmatrix}
	1 & -j \\ 
	j & 1
	\end{bmatrix}
	\begin{bmatrix}
	\cos{(0)} & \cos{(2\pi n/3)} & \cos{(4\pi n/3)}\\
	\sin{(0)} & \sin{(2\pi n/3)} & \sin{(4\pi n/3)}
	\end{bmatrix}}_{\boldsymbol{C}_{\mathrm{H},n}}
\begin{bmatrix}
M_1(s_{+})\\
M_2(s_{+})\\
M_3(s_{+})
\end{bmatrix},\label{eq:MBC_FD_forwMBC}
\end{multline}
where $\boldsymbol{C}_{\mathrm{L},n}$ and $\boldsymbol{C}_{\mathrm{H},n}$ are referred to as the \textit{low} and \textit{high} partial transformation matrices, respectively, due to their association with signals of lower and higher frequencies. By inspection of Eq.~\eqref{eq:MBC_FD_forwMBC} it is already shown that the rotor speed dependent $n$P harmonic is transfered to a DC-component.

\subsection{Reverse MBC transformation}\label{sec:MBC_FreqDomain_Reverse}
Next, the time-domain expression of the reverse MBC transformation is rewritten as
\begin{multline}
\theta_b(t) = \theta_\mathrm{tilt}(t)\left[\cos\left(n\omega_\mathrm{r}t\right)\cos\left(\frac{2\pi n(b-1)}{3}\right)-\sin\left(n\omega_\mathrm{r}t\right)\sin\left(\frac{2\pi n(b-1)}{3}\right)\right]\\
+\theta_\mathrm{yaw}(t)\left[\sin\left(n\omega_\mathrm{r}t\right)\cos\left(\frac{2\pi n(b-1)}{3}\right)+\cos\left(n\omega_\mathrm{r}t\right)\sin\left(\frac{2\pi n(b-1)}{3}\right)\right],
\end{multline}
and is transformed to its frequency-domain representation by
\begin{align}
\nonumber\begin{bmatrix}
\theta_1(s) \\ \theta_2(s) \\ \theta_3(s)
\end{bmatrix} =
\underbrace{\frac{1}{2} 
	\begin{bmatrix}
	\cos{(0)} & \sin{(0)}\\
	\cos{({2\pi n}/{3})} & \sin{({2\pi n}/{3})}\\
	\cos{({4\pi n}/{3})} & \sin{({4\pi n}/{3})}
	\end{bmatrix}
	\begin{bmatrix} 
	1 & -j \\ j & 1 
	\end{bmatrix}}_{\boldsymbol{C}_{\mathrm{L},n}^T}
&\begin{bmatrix} 
\theta_\mathrm{tilt}(s_{-}) \\ \theta_\mathrm{yaw}(s_{-})
\end{bmatrix}\\ &+
\underbrace{\frac{1}{2} 
	\begin{bmatrix}
	\cos{( 0 )} & \sin{( 0 )}\\
	\cos{( {2\pi n}/{3} )} & \sin{( {2\pi n}/{3} )}\\
	\cos{( {4\pi n}/{3} )} & \sin{( {4\pi n}/{3} )}
	\end{bmatrix}
	\begin{bmatrix} 
	1 & j \\ -j & 1 
	\end{bmatrix}}_{\boldsymbol{C}_{\mathrm{H},n}^T}
\begin{bmatrix} 
\theta_\mathrm{tilt}(s_{+}) \\ \theta_\mathrm{yaw}(s_{+})
\end{bmatrix},\label{eq:MBC_FD_revMBC}
\end{align}
where it is seen that the \textit{low} and \textit{high} partial transformation matrices reoccur in a transposed manner. The partial transformation matrices have the remarkable property that $\boldsymbol{C}_{\mathrm{L},n}\boldsymbol{C}_{\mathrm{L},n}^T = 0$ and $\boldsymbol{C}_{\mathrm{H},n}\boldsymbol{C}_{\mathrm{H},n}^T = 0$, which appears to be useful later on.

\subsection{Combining the results: decoupled blade dynamics}\label{sec:MBC_FreqDomain_DRotMod}
\begin{figure}[t!]
	\centering
	\includegraphics[scale=1.0]{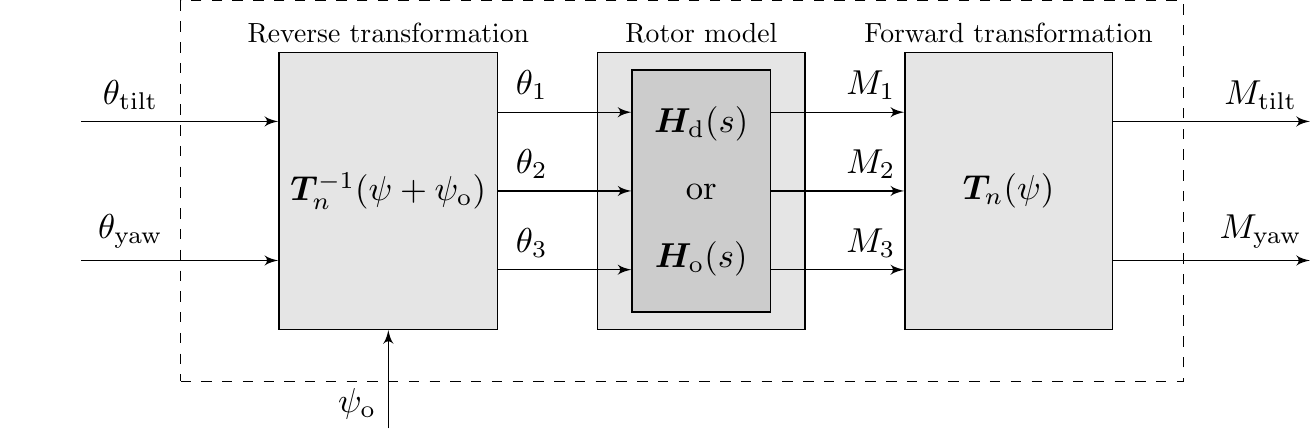}
	\caption{Open-loop non-rotating wind turbine system with fixed-frame input pitch angles $\theta_\mathrm{tilt}$ and $\theta_\mathrm{yaw}$, and output blade moments $M_\mathrm{tilt}$ and $M_\mathrm{yaw}$. For linear analysis purposes, either the diagonal $\boldsymbol{H}_\mathrm{d}(s)$ rotor model including, or the coupled $\boldsymbol{H}_\mathrm{o}(s)$ rotor model excluding cross-terms is considered.}%
	\label{fig:MBC_OpenLoop}%
\end{figure}
Now that the frequency-domain representations of the MBC transformations are defined, the rotor model structure is chosen to be diagonal in this section. In Figure~\ref{fig:MBC_OpenLoop}, the open-loop system with non-rotating pitch angles as input and non-rotating blade moments as output is presented. The diagonal rotor model in the rotating frame is defined as
\begin{align}
\begin{bmatrix}
M_{1}(s)\\
M_{2}(s)\\
M_{3}(s)
\end{bmatrix}
=
\underbrace{\begin{bmatrix}
	H_\mathrm{1}(s) & 0 & 0\\
	0 & H_\mathrm{1}(s) & 0\\
	0 & 0 & H_\mathrm{1}(s)
	\end{bmatrix}}_{\boldsymbol{H}_\mathrm{d}(s)}
\begin{bmatrix}
\theta_{1}(s)\\
\theta_{2}(s)\\
\theta_{3}(s)
\end{bmatrix},\label{eq:MBC_C_WTmodelD}
\end{align}
such that pitch angle $\theta_i(s)$ and blade moment $M_j(s)$ is only related for $i = j$. As will be shown later, the assumption of a diagonal rotor model structure is convenient for analysis purposes, but non-realistic for actual turbines. By substitution of the rotor model from Eq.~\eqref{eq:MBC_C_WTmodelD} into the forward MBC frequency-domain relation in Eq.~\eqref{eq:MBC_FD_forwMBC}, and subsequently substituting Eq.~\eqref{eq:MBC_FD_revMBC}, the following transformed frequency-domain representation is obtained
\begin{multline}
\begin{bmatrix}M_\mathrm{tilt}(s)\\M_\mathrm{yaw}(s)\end{bmatrix} = \frac{2}{3}
\boldsymbol{C}_{\mathrm{L},n} \, H_\mathrm{1}(s_{-}) \, \boldsymbol{I}_{3} 
\left(
{\boldsymbol{C}_{\mathrm{L},n}^T}
\begin{bmatrix} 
\theta_\mathrm{tilt}(s-2jn\omega_\mathrm{r}) \\ \theta_\mathrm{yaw}(s-2jn\omega_\mathrm{r})
\end{bmatrix} +
{\boldsymbol{C}_{\mathrm{H},n}^T}
\begin{bmatrix} 
\theta_\mathrm{tilt}(s) \\ \theta_\mathrm{yaw}(s)
\end{bmatrix} 
\right)\\
+
\frac{2}{3}
\boldsymbol{C}_{\mathrm{H},n} \, H_\mathrm{1}(s_{+}) \, \boldsymbol{I}_{3} 
\left(
{\boldsymbol{C}_{\mathrm{L},n}^T}
\begin{bmatrix} 
\theta_\mathrm{tilt}(s) \\ \theta_\mathrm{yaw}(s)
\end{bmatrix} +
{\boldsymbol{C}_{\mathrm{H},n}^T}
\begin{bmatrix} 
\theta_\mathrm{tilt}(s+2jn\omega_\mathrm{r}) \\ \theta_\mathrm{yaw}(s+2jn\omega_\mathrm{r})
\end{bmatrix} 
\right).
\end{multline}
Since $\boldsymbol{C}_{\mathrm{L},n}\boldsymbol{C}_{\mathrm{L},n}^T = \boldsymbol{C}_{\mathrm{H},n}\boldsymbol{C}_{\mathrm{H},n}^T = 0$, the expression simplifies into
\begin{align}
\begin{bmatrix}M_\mathrm{tilt}(s)\\M_\mathrm{yaw}(s)\end{bmatrix} = 
\frac{1}{2}
\left(
H_\mathrm{1}(s_{-}) \, \boldsymbol{I}_{2} 
\begin{bmatrix}
1 & j \\ -j & 1
\end{bmatrix} +
H_\mathrm{1}(s_{+}) \, \boldsymbol{I}_{2} 
\begin{bmatrix}
1 & -j \\ j & 1
\end{bmatrix}
\right)
\begin{bmatrix} 
\theta_\mathrm{tilt}(s) \\ \theta_\mathrm{yaw}(s)
\end{bmatrix},\label{eq:MBC_C_TFexpression}
\end{align}
where $\boldsymbol{I}_\mathrm{2}\in \mathbb{R}^{2\times2}$ is an identity matrix, and is rewritten as the transfer function matrix
\begin{align}
\begin{bmatrix}M_\mathrm{tilt}(s)\\M_\mathrm{yaw}(s)\end{bmatrix} &= 
\underbrace{
	\frac{1}{2}\begin{bmatrix}
	{H_\mathrm{1}(s_{-})+H_\mathrm{1}(s_{+})}  & j{H_\mathrm{1}(s_{-})-jH_\mathrm{1}(s_{+})}\\
	-j{H_\mathrm{1}(s_{-})+jH_\mathrm{1}(s_{+})} & {H_\mathrm{1}(s_{-})+H_\mathrm{1}(s_{+})}
	\end{bmatrix}
}_{\boldsymbol{P}_\mathrm{d}(s,\omega_\mathrm{r})}
\begin{bmatrix} 
\theta_\mathrm{tilt}(s) \\ \theta_\mathrm{yaw}(s)
\end{bmatrix}.\label{eq:MBC_C_TFmatrixD}
\end{align}
Although the wind turbine blade models $H_\mathrm{1}(s)$ in Eq.~\eqref{eq:MBC_C_WTmodelD} are implemented in a decoupled way, it is seen that the off-diagonal terms are non-zero when the response of $H(s)$ is frequency dependent (non-constant). Thus, the presumably decoupled tilt and yaw-axes show cross-coupling in ${\boldsymbol{P}_\mathrm{d}(s,\omega_\mathrm{r})}$ when a diagonal and dynamic rotor model is considered. This conclusion was drawn earlier \citep{lu2015analysis}. However, in the next section, the assumption of a diagonal rotor model is alleviated by the introduction of cross-terms.

\subsection{Combining the results: coupled blade dynamics}\label{sec:MBC_FreqDomain_CRotMod}
In the previous section, the rotor model was assumed to consist of decoupled blade models. Now, this assumption is alleviated by incorporating off-diagonal blade models
\begin{align}
\begin{bmatrix}
M_{1}(s)\\
M_{2}(s)\\
M_{3}(s)
\end{bmatrix}
=
\underbrace{\begin{bmatrix}
	H_\mathrm{1}(s) & H_\mathrm{2}(s) & H_\mathrm{2}(s)\\
	H_\mathrm{2}(s) & H_\mathrm{1}(s) & H_\mathrm{2}(s)\\
	H_\mathrm{2}(s) & H_\mathrm{2}(s) & H_\mathrm{1}(s)
	\end{bmatrix}}_{\boldsymbol{H}_\mathrm{o}(s)}
\begin{bmatrix}
\theta_{1}(s)\\
\theta_{2}(s)\\
\theta_{3}(s)
\end{bmatrix},\label{eq:MBC_C_WTmodelO}
\end{align}
such that coupling is also present between pitch angle $\theta_i(s)$ and blade moment $M_j(s)$ for $i \neq j$ by $H_\mathrm{2}(s)$: in Section~\ref{sec:NLA_LinSpectral} it is shown that this model structure represents the interactions of high-fidelity model linearizations. The derivation to arrive at the transfer function matrix $\boldsymbol{P}_\mathrm{o}(s,\omega_\mathrm{r})$ is omitted in this section, as it follows a similar procedure given in the previous section. The resulting matrix is given by
\begin{align}
\begin{bmatrix}M_\mathrm{tilt}(s)\\M_\mathrm{yaw}(s)\end{bmatrix} &= 
\underbrace{
	\frac{1}{2}\begin{bmatrix}
	H_\mathrm{12}(s_{-})+H_\mathrm{12}(s_{+}) & jH_\mathrm{12}(s_{-})-jH_\mathrm{12}(s_{+})\\
	-jH_\mathrm{12}(s_{-})+jH_\mathrm{12}(s_{+}) & H_\mathrm{12}(s_{-})+H_\mathrm{12}(s_{+})
	\end{bmatrix}
}_{\boldsymbol{P}_\mathrm{o}(s,\omega_\mathrm{r})}
\begin{bmatrix} 
\theta_\mathrm{tilt}(s) \\ \theta_\mathrm{yaw}(s)
\end{bmatrix},\label{eq:MBC_C_TFmatrixO}
\end{align}
with
\begin{align}
H_\mathrm{12}(s) &= {H_\mathrm{1}(s)-H_\mathrm{2}(s)}.\label{eq:MBC_C_TFH12}
\end{align}
As will be shown later, the obtained model structure is better able to identify the optimal azimuth offset opposed to the result from Section~\ref{sec:MBC_FreqDomain_DRotMod}, for operating conditions with increased dynamic blade coupling. The following section incorporates the azimuth offset in the framework for both the decoupled and coupled rotor model structures.

\subsection{Inclusion of the azimuth offset}\label{sec:MBC_AziOffset}
In this section, the effect on the main and off-diagonal terms by incorporating an azimuth offset  $\psi_\mathrm{o}\in\mathbb{R}$ in the reverse transformation is considered: variables subject to the effect of the offset are denoted with a tilde $\tilde{\left(\cdot\right)}$. Multiplication of the transformation matrices $\boldsymbol{T}(\psi) \tilde{\boldsymbol{T}}^{-1}(\psi+ \psi_\mathrm{o})$ for $\psi_\mathrm{o}\neq0$ does not result in an identity matrix, and influences the diagonal and off-diagonal terms in the transfer function matrix $\tilde{\boldsymbol{P}}$. For evaluation of this effect, Eq.~\eqref{eq:MBC_FD_revMBC} is expanded by adding the azimuth offset to the nominal azimuth such that the following expression is obtained
\begin{multline}\label{eq:thetab_s_offset}
\begin{bmatrix}
\theta_1(s) \\ \theta_2(s) \\ \theta_3(s)
\end{bmatrix} =
\underbrace{\frac{1}{2}
	\begin{bmatrix}
	\cos{(n\psi_\mathrm{o})} & \sin{(n\psi_\mathrm{o})}\\
	\cos{(n({2\pi}/{3}+\psi_\mathrm{o}))} & \sin{(n({2\pi}/{3}+\psi_\mathrm{o}))}\\
	\cos{(n({4\pi}/{3}+\psi_\mathrm{o}))} & \sin{(n({4\pi}/{3}+\psi_\mathrm{o}))}
	\end{bmatrix}
	\begin{bmatrix} 
	1 & -j \\ j & 1 
	\end{bmatrix}}_{\tilde{\boldsymbol{C}}_{\mathrm{L},n}^T(\psi_\mathrm{o})}
\begin{bmatrix} 
\theta_\mathrm{tilt}(s_{-}) \\ \theta_\mathrm{yaw}(s_{-})
\end{bmatrix} \\
+ \underbrace{\frac{1}{2} 
	\begin{bmatrix}
	\cos{(n\psi_\mathrm{o})} & \sin{(n\psi_\mathrm{o})}\\
	\cos{(n({2\pi}/{3}+\psi_\mathrm{o}))} & \sin{(n({2\pi}/{3}+\psi_\mathrm{o}))}\\
	\cos{(n({4\pi}/{3}+\psi_\mathrm{o}))} & \sin{(n({4\pi}/{3}+\psi_\mathrm{o}))}
	\end{bmatrix}
	\begin{bmatrix} 
	1 & j \\ -j & 1 
	\end{bmatrix}}_{\tilde{\boldsymbol{C}}_{\mathrm{H},n}^T(\psi_\mathrm{o})}
\begin{bmatrix} 
\theta_\mathrm{tilt}(s_{+}) \\ \theta_\mathrm{yaw}(s_{+})
\end{bmatrix}. 
\end{multline}
where the partial transformation matrices now include the azimuth offset and are redefined using trigonometric identities as
\begin{align}
\tilde{\boldsymbol{C}}_{\mathrm{L},n}^T(\psi_\mathrm{o}) &= \frac{1}{2} 
\begin{bmatrix}
\cos{(0)} & \sin{(0)}\\
\cos{({2\pi n}/{3})} & \sin{({2\pi n}/{3})}\\
\cos{({4\pi n}/{3})} & \sin{({4\pi n}/{3})}
\end{bmatrix} 
\begin{bmatrix}
\cos{(n\psi_\text{o})}  & \sin{(n\psi_\text{o})}\\
-\sin{(n\psi_\text{o})} & \cos{(n\psi_\text{o})}
\end{bmatrix}
\begin{bmatrix} 
1 & -j \\ j & 1 
\end{bmatrix},\\
\tilde{\boldsymbol{C}}_{\mathrm{H},n}^T(\psi_\mathrm{o}) &= \frac{1}{2} 
\begin{bmatrix}
\cos{( 0 )} & \sin{( 0 )}\\
\cos{( {2\pi n}/{3} )} & \sin{( {2\pi n}/{3} )}\\
\cos{( {4\pi n}/{3} )} & \sin{( {4\pi n}/{3} )}
\end{bmatrix} 
\begin{bmatrix}
\cos{(n\psi_\text{o})}  & \sin{(n\psi_\text{o})}\\
-\sin{(n\psi_\text{o})} & \cos{(n\psi_\text{o})}
\end{bmatrix}
\begin{bmatrix} 
1 & j \\ -j & 1 
\end{bmatrix}.
\end{align}
Comparing the partial transformation matrices to the results obtained earlier in Eqs.\eqref{eq:MBC_FD_forwMBC}~and~\eqref{eq:MBC_FD_revMBC} shows the addition of a rotation matrix. By applying the correct (optimal) phase offset, the rotation matrix corrects for the phase losses in the rotating frame, and lets the transformed axes coincide with the vertical tilt and horizontal yaw axes in the non-rotating frame. Furthermore, the matrix is a normalized version of the steady-state gain matrix of the inverse plant \cite{ref:Unguran2019FFIPC}, and can alternatively be taken as part of the controller outside the transformed system.

By deriving the transformation matrix for the decoupled rotor model structure, now including the azimuth offset, results in
\begin{align}\label{eq:MBC_AO_TFmatrixD}
\begin{bmatrix}M_\text{tilt}(s)\\M_\text{yaw}(s)\end{bmatrix} = 
\underbrace{
	\frac{1}{2}\begin{bmatrix}
	{H(s_{-})\tilde{p}(\psi_\mathrm{o})+H(s_{+})\tilde{q}(\psi_\mathrm{o})} & j{H(s_{-})\tilde{p}(\psi_\mathrm{o})-jH(s_{+})\tilde{q}(\psi_\mathrm{o})}\\
	-j{H(s_{-})\tilde{p}(\psi_\mathrm{o})+jH(s_{+})\tilde{q}(\psi_\mathrm{o})} & {H(s_{-})\tilde{p}(\psi_\mathrm{o})+H(s_{+})\tilde{q}(\psi_\mathrm{o})}
	\end{bmatrix}
}_{\tilde{\boldsymbol{P}}_\mathrm{d}(s,\omega_\mathrm{r},\psi_\mathrm{o})}
\begin{bmatrix}
\theta_\text{tilt}(s) \\ \theta_\text{yaw}(s)
\end{bmatrix},
\end{align}
whereas the matrix is defined for the coupled case as
\begin{align}\label{eq:MBC_AO_TFmatrixO}
\begin{bmatrix}M_\text{tilt}(s)\\M_\text{yaw}(s)\end{bmatrix} = 
\underbrace{
	\frac{1}{2}\begin{bmatrix}
	{H_\mathrm{12}(s_{-})\tilde{p}(\psi_\mathrm{o})+H_\mathrm{12}(s_{+})\tilde{q}(\psi_\mathrm{o})} & j{H_\mathrm{12}(s_{-})\tilde{p}(\psi_\mathrm{o})-jH_\mathrm{12}(s_{+})\tilde{q}(\psi_\mathrm{o})}\\
	-j{H_\mathrm{12}(s_{-})\tilde{p}(\psi_\mathrm{o})+jH_\mathrm{12}(s_{+})\tilde{q}(\psi_\mathrm{o})} & {H_\mathrm{12}(s_{-})\tilde{p}(\psi_\mathrm{o})+H_\mathrm{12}(s_{+})\tilde{q}(\psi_\mathrm{o})}
	\end{bmatrix}
}_{\tilde{\boldsymbol{P}}_\mathrm{o}(s,\omega_\mathrm{r},\psi_\mathrm{o})}
\begin{bmatrix}
\theta_\text{tilt}(s) \\ \theta_\text{yaw}(s)
\end{bmatrix},
\end{align}
where $\tilde{p}(\psi_\mathrm{o})$ and $\tilde{q}(\psi_\mathrm{o})$ are
\begin{align}
\tilde{p}(\psi_\mathrm{o}) &= \cos{(n\psi_\mathrm{o})} - j\sin{(n\psi_\mathrm{o})},\\
\tilde{q}(\psi_\mathrm{o}) &= \cos{(n\psi_\mathrm{o})} + j\sin{(n\psi_\mathrm{o})}.
\end{align}
From the above derived result it is concluded that the azimuth offset influences the main and off-diagonal terms for both the coupled and decoupled cases. By comparing Eqs.~\ref{eq:MBC_AO_TFmatrixD}~and~\ref{eq:MBC_AO_TFmatrixO}, it is observed that both are similar, but the latter mentioned differs in a way that cross-coupling between the blade models influences the non-rotating dynamics. As a result, the optimal offset value will be different for both cases. An analysis using simplified blade models is given in the next section.

\section{Analysis on simplified rotor models}\label{sec:LA}
This section showcases the effect and implications of the azimuth offset using simplified models, for both decoupled and coupled rotor model structures in Sections~\ref{sec:LA_Decoupled}~and~\ref{sec:LA_Coupled}, respectively. First-order linear dynamic blade models are taken, as this allows for a convenient assessment of the offset effects: application of higher-order models would result in a similar analysis.

\subsection{Decoupled blade dynamics}\label{sec:LA_Decoupled}
\begin{figure}[b!]
	\centering
	\includegraphics[scale=1.0]{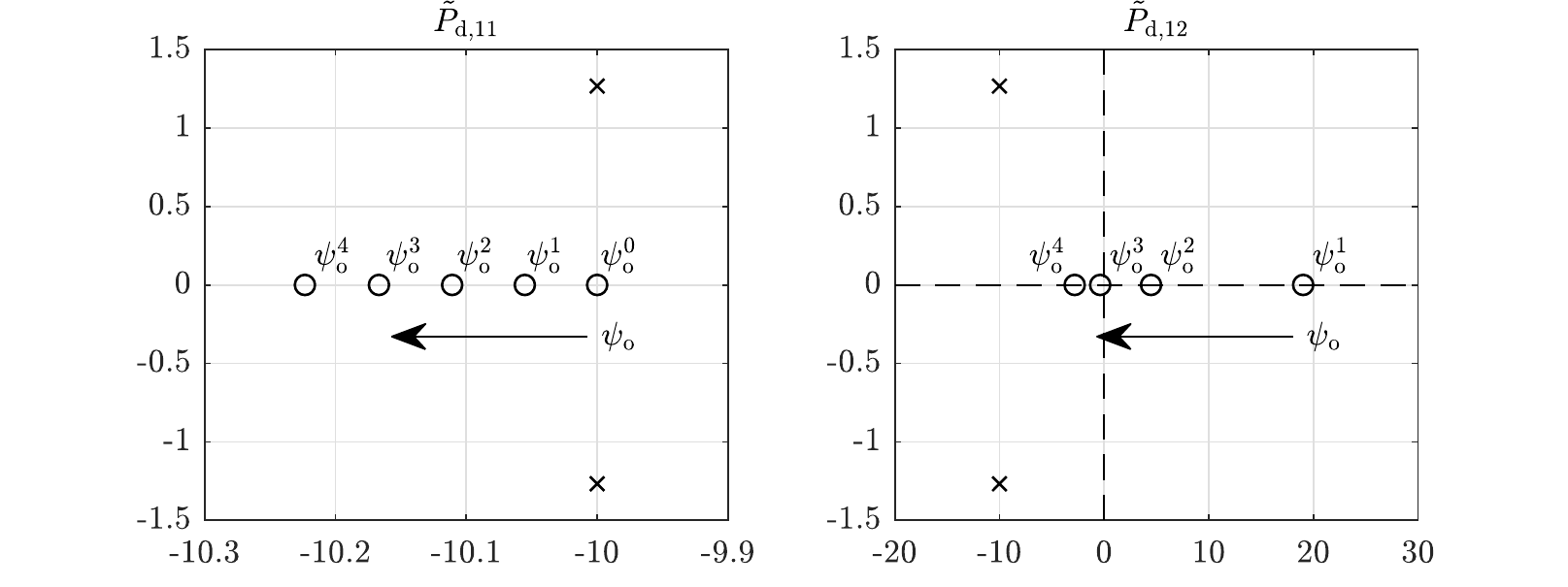}
	\caption{Pole-zero map for the main- and off-diagonal transfer functions in $\tilde{P}_{11}$ and $\tilde{P}_{12}$, respectively. It is shown that for the assumed model $H_\mathrm{1}(s)$ with $K_1=1$ and $\tau_1=0.1$, the azimuth offset influences the location of the open-loop zeros ($\circ$) in both cases; the pole ($\times$) locations remain unchanged. The magnitude of the cross-terms is minimized by choosing the optimal offset value $\psi^*_\mathrm{o}$.}%
	\label{fig:LA_pzmap}%
\end{figure}
The decoupled rotor model is made up of first-order blade models of the form
\begin{align}
H_1(s) &= \frac{M_b}{\theta_b} = K_1\frac{1}{\tau_1 s + 1},\label{eq:LA_H1}
\end{align}
where $K_1$ is the steady-state gain and $\tau_1$ the time constant of the transfer function. As the main-diagonal elements of $\tilde{\boldsymbol{P}}_\mathrm{d}$ are equal and the off-diagonal elements are the same up to a sign-change, only the transfer functions in the matrix upper row are considered. By substitution of $s=j\omega$, the frequency response function of the diagonal elements is given by
\begin{align}
\tilde{P}_{\mathrm{d},11}(j\omega,\omega_\mathrm{r},\psi_\mathrm{o}) &= \tilde{P}_{\mathrm{d},22}(j\omega,\omega_\mathrm{r},\psi_\mathrm{o}) = K_1\frac{\tau_1\omega\cos\left(\psi_\mathrm{o} \right)-(\tau_1\mathrm{\omega_\mathrm{r}}\sin\left(\psi_\mathrm{o} \right)+\cos\left(\psi_\mathrm{o} \right))j}{2\tau_1\omega+(\tau_1^2\omega^2-\tau_1^2\omega_\mathrm{r}^2-1)j},\label{eq:LA_PTildeOn}
\end{align}
and the frequency response functions of the off-diagonal terms are represented by
\begin{align}
\tilde{P}_{\mathrm{d},12}(j\omega,\omega_\mathrm{r},\psi_\mathrm{o}) &= -\tilde{P}_{\mathrm{d},21}(j\omega,\omega_\mathrm{r},\psi_\mathrm{o}) = K_1\frac{-\tau_1\omega\sin{(\psi_\mathrm{o})}-\left(\tau_1\omega_\mathrm{r}\cos{(\psi_\mathrm{o})}-\sin{(\psi_\mathrm{o})}\right)j}{2\tau_1\omega+(\tau_1^2\omega^2-\tau_1^2\omega_\mathrm{r}^2-1)j}.\label{eq:LA_PTildeOff}
\end{align}
In both expressions the azimuth offset only occurs in the numerator. For the off-diagonal expression in Eq.~\eqref{eq:LA_PTildeOff}, the low-frequency magnitude ($\omega\to 0$) can be attenuated using the offset. In effect, the complex term in the frequency response function of Eq.~\eqref{eq:LA_PTildeOff} cancels out, and minimizes the low-frequency gain.

\begin{figure}[t!]
	\centering
	\includegraphics[scale=1.0]{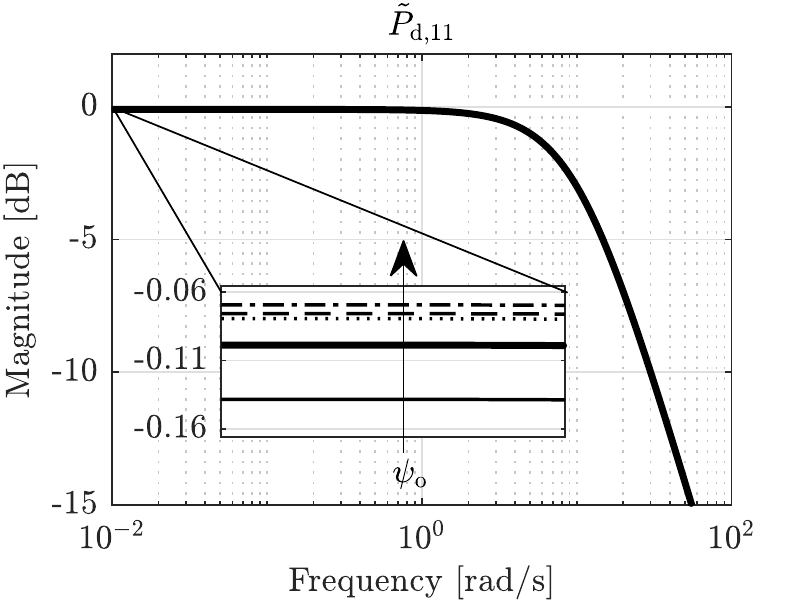}
	\includegraphics[scale=1.0]{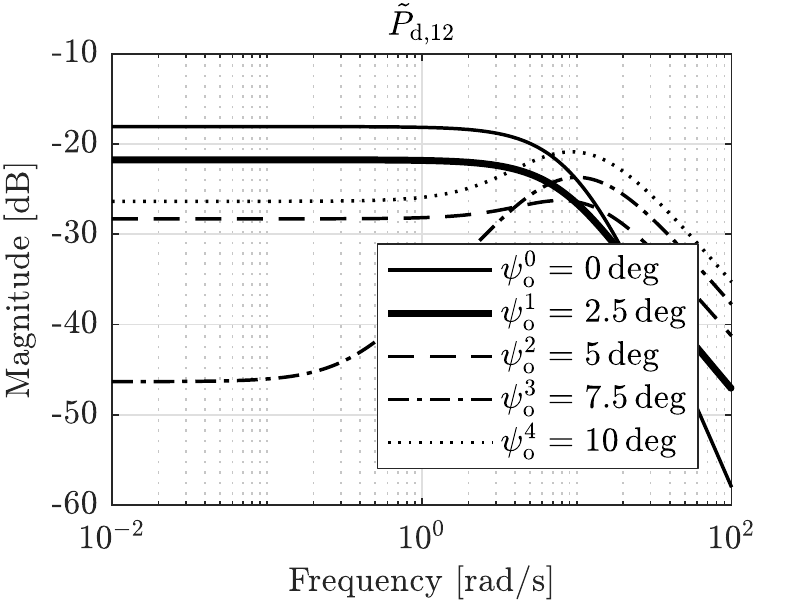}
	\caption{Bode diagrams of $\tilde{P}_{\mathrm{d},11}$ and $\tilde{P}_{\mathrm{d},12}$ for different $\psi_\mathrm{o}$. The steady-state gain of the diagonal term increases, whereas the gain of the off-diagonal term decreases up to a certain offset value.}%
	\label{fig:LA_bode}%
\end{figure}

For illustration purposes, the transfer function $H_\mathrm{1}(s)$ is taken with a steady-state gain $K_1 = 1$, a time constant $\tau_1 = 0.1$~s and a rotor speed $\omega_\mathrm{r} = 1.27$~rad\,s\textsuperscript{-1}, which is the rated speed of the NREL~5\nobreakdash-MW reference turbine. In Figure~\ref{fig:LA_pzmap}, pole-zero diagrams are given for the transfer function elements $\tilde{P}_\mathrm{d,11}$ and $\tilde{P}_\mathrm{d,12}$. For the latter mentioned transfer function, the offset introduces a zero which is non-present in the case of $\psi_\mathrm{o} = 0$. The offset is used to actively influence the zero location, and does not affect the pole locations. The zero attains a lower real value for increasing offsets. The optimal offset moves the introduced off-diagonal zero to the imaginary axis to form a pure differentiator, of which the effect is shown in Figure~\ref{fig:LA_bode}.  For the same optimal offset, the steady-state gain of the diagonal term is maximized. The influence of the offset on the main-diagonal steady-state low-frequency gain should be taken into account during controller design. That is, including the optimal offset increases the bandwidth of the open-loop gain.

\begin{figure}[t!]
	\centering
	\includegraphics[scale=1.0]{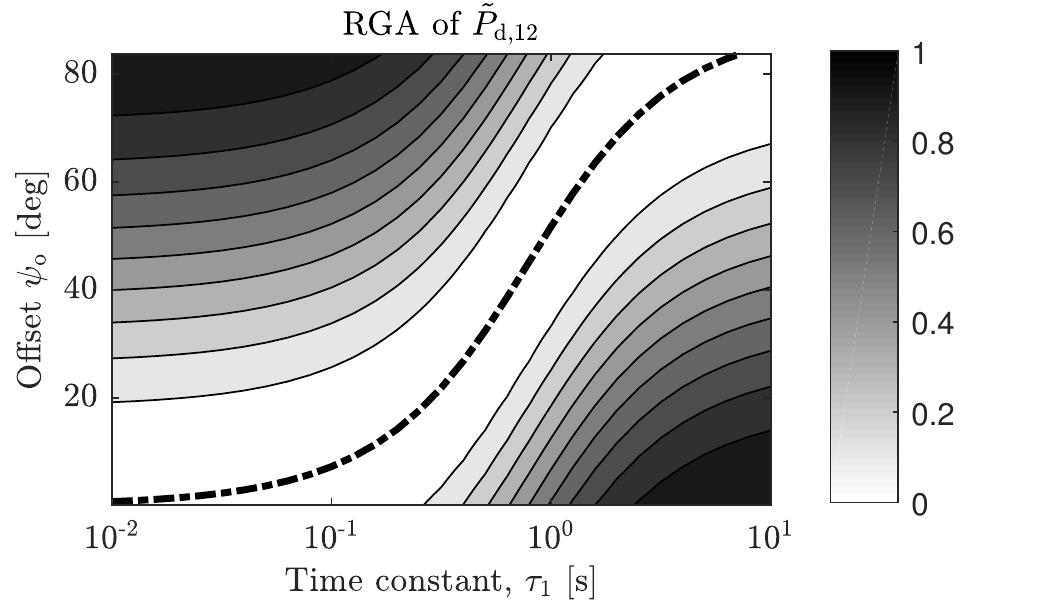}
	\caption{RGA of $\tilde{P}_{\mathrm{d},12}$ evaluated at $\omega = 0$ for the decoupled rotor model structure. The dash-dotted line represents the optimal offset found by the analytical expression. It is shown that the optimal offset is highly dependent on the model dynamics.}%
	\label{fig:RGA_12Diag}%
\end{figure}

For a decoupled rotor model consisting of first-order blade dynamics, the optimal offset is analytically computed by
\begin{align}
\psi_\mathrm{o,d}^{*} = \tan^{-1}{(\tau_1\omega_\mathrm{r})}\label{eq:LA_Decoupled_psi_od}.
\end{align}
Calculation of the optimal offset results in $\psi_\mathrm{o,d}^*=7.22$~deg, which is in accordance to the near-optimal result found in Figure~\ref{fig:LA_bode}. Figure~\ref{fig:RGA_12Diag} presents the RGA of $\tilde{P}_\mathrm{d,12}$ over a range of first-order model time constants and azimuth offsets. It is shown that a clear optimal offset path is present, which is predicted using the analytic expression given above. It is furthermore concluded that for the decoupled blade model case, the optimal offset is equal to the phase loss of the blade pitch to blade moment system at the considered $n$P harmonic. Eq.~\eqref{eq:LA_Decoupled_psi_od} also shows that the optimal offset is dependent on the rotor speed, which is of importance when IPC is applied in the below-rated operating region.

\subsection{Coupled blade dynamics}\label{sec:LA_Coupled}
\begin{figure}[b!]
	\centering
	\includegraphics[scale=1.0]{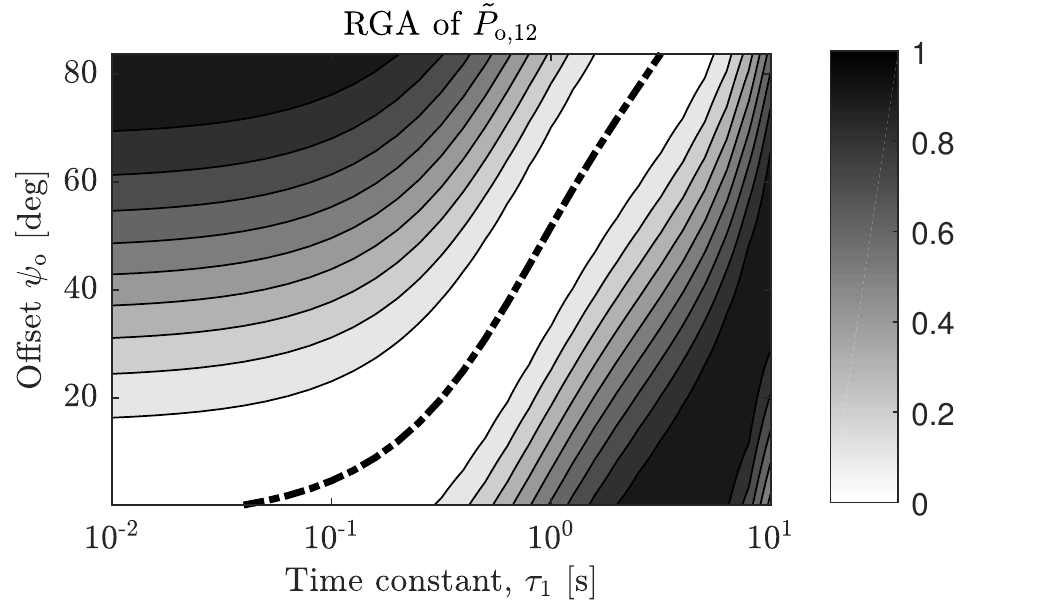}
	\caption{RGA of $\tilde{P}_{\mathrm{o},12}$ evaluated at $\omega = 0$ for the coupled rotor model structure. The dash-dotted line represents the optimal offset found by the analytical expression. It is shown that the optimal offset is highly dependent on the combined diagonal and off-diagonal dynamic model characteristics and differs significantly from the characteristics found for the decoupled case.}%
	\label{fig:RGA_12OffDiag}%
\end{figure}
The derivation is now performed for the rotor model with coupled blade dynamics, $\tilde{\boldsymbol{P}}_\mathrm{o}$. The main-diagonal transfer function $H_1(s)$ is taken as in Eq.~\eqref{eq:LA_H1}, whereas two distinct cases for the off-diagonal model $H_2(s)$ are examined. The first case is a reduced magnitude version of $H_1(s)$ with $K_2 = \delta K_1$ where $\left\{\delta \subset\mathbb{R}~|~0<\delta<1 \right\}$, and the second case additionally has a time constant $\tau_2\neq\tau_1$. The transfer function is given by
\begin{align}
H_2(s) &= \frac{M_i}{\theta_j} = K_2\frac{1}{\tau_2 s+1}\quad \text{with}\quad i\neq j,
\end{align}
and according to Eq.~\eqref{eq:MBC_C_TFH12}, the resulting expressions of the combined transfer functions become
\begin{align}
&\text{\bf{Case 1}: $K_2 = \delta K_1,~\tau_1=\tau_2$}\qquad H_{12}^1(s) = K_1(1-\delta)\frac{1}{\tau_1 s + 1}\label{eq:LA_Coupled_K},\\
&\text{\bf{Case 2}: $K_2 = \delta K_1,~\tau_1\neq\tau_2$}\qquad H_{12}^2(s) = \frac{K_1(\tau_2s+1)-K_2(\tau_1s+1)}{(\tau_1s+1)(\tau_2s+1)}.\label{eq:LA_Coupled_Ktau}
\end{align}
By comparing Eq.~\eqref{eq:LA_H1}~and~\eqref{eq:LA_Coupled_K} it is immediately recognized that for the first case, the result is only scaled by a factor $\delta$ and does not influence the optimal offset. However, for the second case, the resulting transfer function changes significantly for which the derivation is performed. The resulting elements of the matrix upper row of $\tilde{\boldsymbol{P}}_\mathrm{o}$ are
\begin{align}
\tilde{P}_{\mathrm{o},11}&(j\omega,\omega_\mathrm{r},\psi_\mathrm{o}) = \tilde{P}_{\mathrm{o},22}(j\omega,\omega_\mathrm{r},\psi_\mathrm{o})\nonumber \\
&=\frac{\tilde{p}(\psi_\mathrm{o})}{2}\left(\frac{K_{1}}{\tau _{1}\left(\omega-\omega_\mathrm{r}\right)j+1}-\frac{K_{2}}{\tau _{2}\left(\omega-\omega_\mathrm{r}\right)j+1}\right)+\frac{\tilde{q}(\psi_\mathrm{o})}{2}\left(\frac{K_{1}}{\tau _{1}\left(\omega+\omega_\mathrm{r}\right)j+1}-\frac{K_{2}}{\tau _{2}\left(\omega+\omega_\mathrm{r}\right)j+1}\right),\\
\tilde{P}_{\mathrm{o},12}&(j\omega,\omega_\mathrm{r},\psi_\mathrm{o}) = -\tilde{P}_{\mathrm{o},21}(j\omega,\omega_\mathrm{r},\psi_\mathrm{o})\nonumber \\
&=\frac{\tilde{p}(\psi_\mathrm{o})j}{2}\left(\frac{K_{1}}{\tau _{1}\left(\omega-\omega_\mathrm{r}\right)j+1}-\frac{K_{2}}{\tau _{2}\left(\omega-\omega_\mathrm{r}\right)j+1}\right)-\frac{\tilde{q}(\psi_\mathrm{o})j}{2}\left(\frac{K_{1}}{\tau _{1}\left(\omega+\omega_\mathrm{r}\right)j+1}-\frac{K_{2}}{\tau _{2}\left(\omega+\omega_\mathrm{r}\right)j+1}\right).
\end{align}
Further substitution and manipulations of the above given relations lead to cumbersome expressions. However, also in this case it is possible to nullify the numerator using the optimal azimuth offset given by the analytic expression
\begin{align}
\psi_\mathrm{o,o}^* &= \tan^{-1}{\left( \frac{K_1\tau_1(1+\tau_2^2\omega_\mathrm{r}^2) - K_2\tau_2(1+\tau_1^2\omega_\mathrm{r}^2)}{K_1(1+\tau_2^2\omega_\mathrm{r}^2)-K_2(1+\tau_1^2\omega_\mathrm{r}^2)}\omega_\mathrm{r} \right)},\label{eq:LA_Decoupled_psi_oo}
\end{align} 
where for the case $K_2 = 0$ (no coupling), the relation reduces to the expression given by Eq.~\eqref{eq:LA_Decoupled_psi_od}.
\begin{figure}[b!]
	\centering
	\includegraphics[scale=1.0]{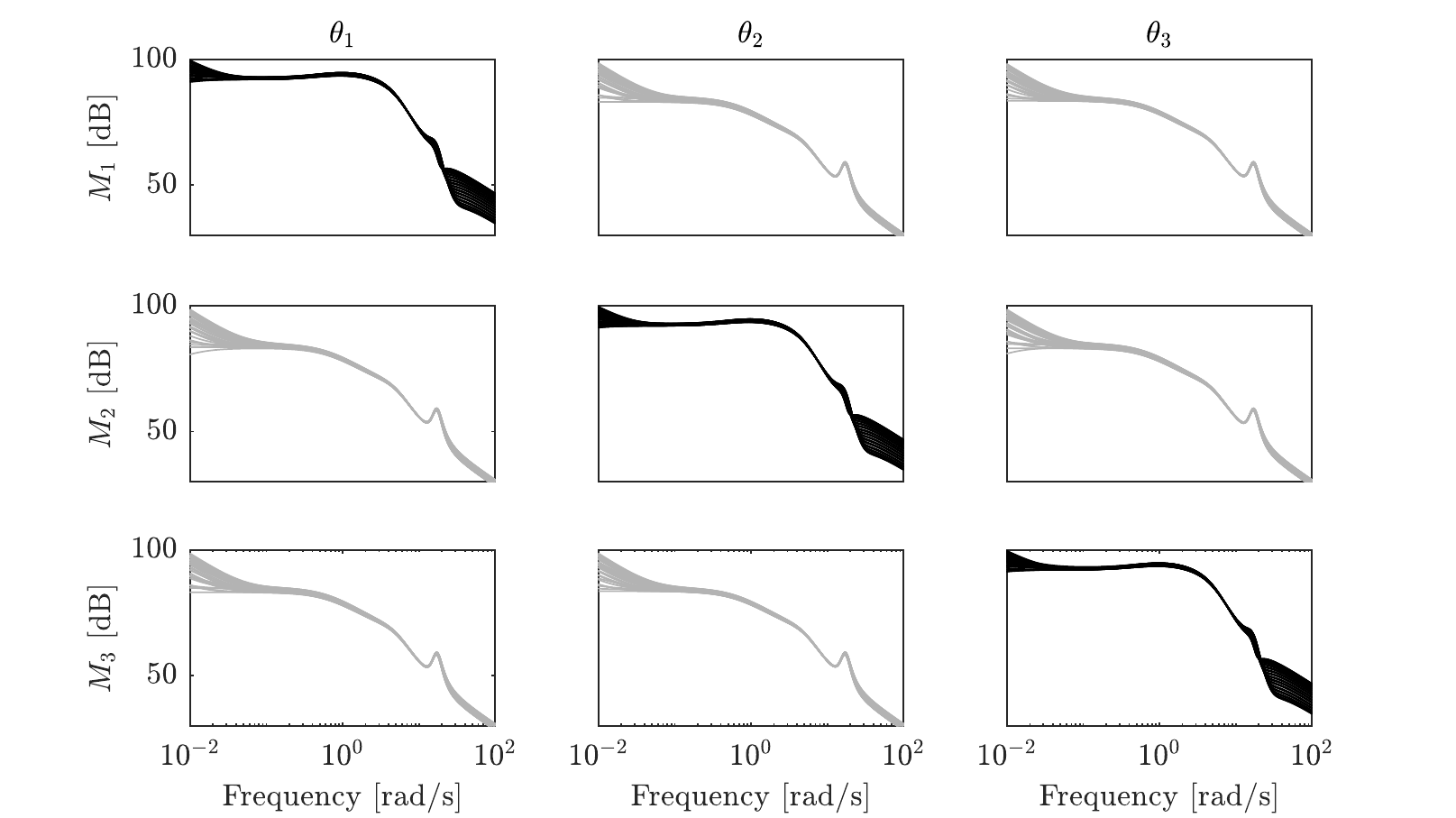}
	\caption{Main- and off-diagonal linear models of the NREL~5\nobreakdash-MW blade dynamics in black and gray, respectively, showing the dynamics from blade pitch $\theta_i$ to out-of-plane blade root moment $M_j$ in the rotating frame. It is shown that the off-diagonal dynamics have an overall reduced, but non-negligible magnitude compared to the main-diagonal elements.}%
	\label{fig:FreqRespLinearModel}%
\end{figure}

For illustration purposes, the constants $K_1$, $\tau_1$ and $\omega_\mathrm{r}$ are taken as in Section~\ref{sec:LA_Decoupled}, and $K_2 = 0.1$ and $\tau_2=1$~s. Using these values, the optimal offset is calculated being $\psi_\mathrm{o,o}^* = 4.60$~deg, which differs from the result found in the previous section. Furthermore, Figure~\ref{fig:RGA_12OffDiag} shows the off-diagonal RGA for the coupled rotor case. It is shown that the decoupling characteristics differ significantly from the results obtained in Figure~\ref{fig:RGA_12Diag}, especially for higher time constants (slower blade dynamics). The main conclusion of this section is that the chosen rotor model structure, including or excluding blade dynamic coupling, has a high influence on the analysis for finding the optimal offset value.

\section{Results on the NREL~5\nobreakdash-MW reference wind turbine}\label{sec:NLA}
The previous section shows significant improvements on the decoupling of transformed model structures using simplified blade models. This section is devoted to the validation of the described theory on linearizations of the NREL~5\nobreakdash-MW reference wind turbine. In Section~\ref{sec:NLA_LinSpectral}, linearizations of the NREL~5\nobreakdash-MW reference turbine are obtained and used in Section~\ref{sec:NLA_LinSpectralOffset} to compute the optimal offset. The results are subsequently validated against the non-parametric spectral models presented in the problem formalization (Section~\ref{sec:MBC_ProblemStatement}).

\subsection{Obtaining linearizations in the rotating frame}\label{sec:NLA_LinSpectral}
Linearizations of the NREL~5\nobreakdash-MW turbine are obtained using an extension \citep{ref:FASTv8GUI} for NREL's FAST v8.16. The extension program includes a Graphical User Interface (GUI) and functionality for determining trim conditions prior to the open-loop simulations for linearization. Linear models are obtained for wind speeds $U = 5-25$~m\,s\textsuperscript{-1}.

The resulting state-space model for each wind speed consists of the system $A\in\mathbb{R}^{r\times r\times k}$, input $B\in\mathbb{R}^{r\times p\times k}$, output $C\in\mathbb{R}^{q\times r\times k}$ and direct feedthrough $D\in\mathbb{R}^{q\times p\times k}$ matrices. Over a full rotor rotation, $k = 36$ evenly spaced models are obtained with a model order $r=14$ and $p=q=3$ in- and outputs. Figure~\ref{fig:FreqRespLinearModel} presents the linearization results by means of Bode magnitude plots from blade pitch to blade moment for a wind speed of $U = 25$~m\,s\textsuperscript{-1}. This wind speed is chosen as an exemplary case, as the effect of dynamic blade coupling becomes more apparent for higher wind speed conditions. As the models are defined in a rotating reference frame, the dynamics vary with the rotor position. However, it can be seen that the dynamics from $\theta_i$ to $M_j$ show similar dynamics for both $i=j$ and $i\neq j$. The linearizations include first-order pitch actuator dynamics with a bandwidth of $\omega_\mathrm{a} = 2.5$~rad\,s\textsuperscript{-1}. The next sections elaborate on the effect of including and excluding the cross terms in the analysis.

\subsection{Transforming linear models and evaluating decoupling}\label{sec:NLA_LinSpectralOffset}
As recognized previously by inspection of Figure~\ref{fig:FreqRespLinearModel}, the set of diagonal and off-diagonal models show similar dynamics. The effect of this coupling on the optimal azimuth offset is investigated in this section using linearizations of the NREL~5\nobreakdash-MW turbine.
\begin{figure}[b!]
	\centering
	\includegraphics[scale=1.0]{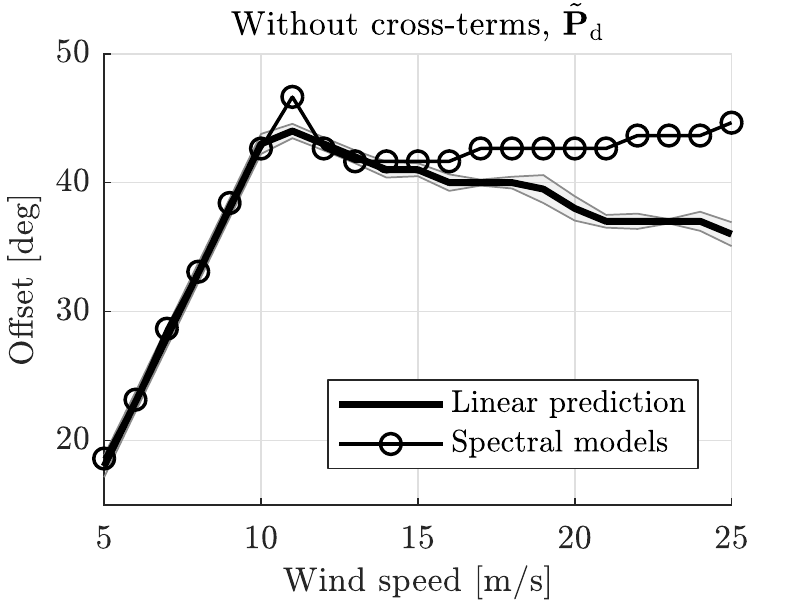}
	\includegraphics[scale=1.0]{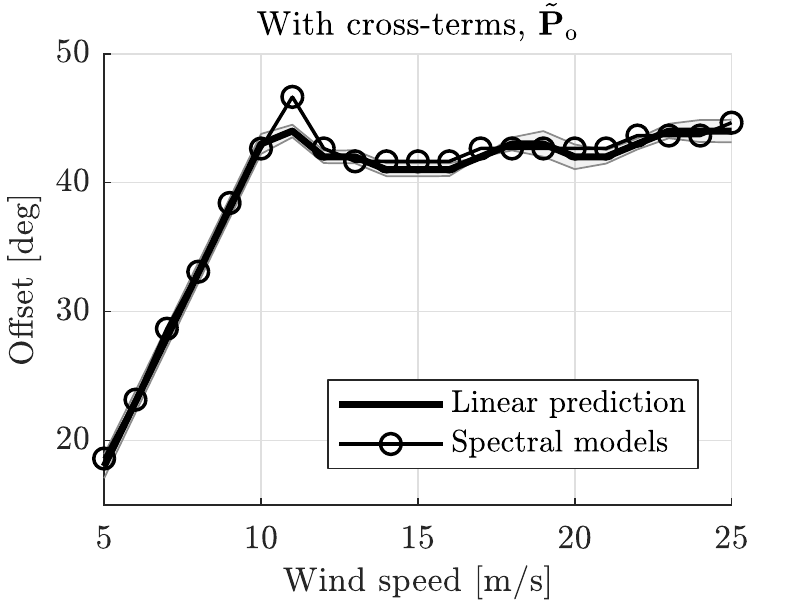}
	\caption{Linear prediction of the optimal azimuth offset over $k$ linearizations, where the median per wind speed is taken as the optimal offset value. The transformation is applied for the cases of decoupled (left) and coupled (right) blade dynamics. It is shown that the inclusion of blade coupling is able to better explain the results obtained from spectral analysis.}
	\label{fig:R_BoxPlotAziOff}%
\end{figure}

Up to this point, the analysis of the effect of the azimuth offset is illustrated using a Multiple-Input Multiple-Output (MIMO) transfer function representation. However, transforming higher-order models (e.g., linearizations obtained from FAST) in this representation can become numerically challenging. Therefore, Appendix~\ref{sec:AppA} includes a derivation of the MBC transformation including the offset in the state-space system representation. Because this approach only requires subsequent matrix multiplications, the implementation is faster and numerically more stable. However, in the remainder of this paper, the transfer function representation is used to highlight insights for various problem aspects.

In this section, by using the transfer function representation, the off-diagonal elements are easily included and excluded from the analysis. Therefore, the obtained linear state-space systems are converted to transfer functions and transformed to symbolic expressions for substitution of the Laplace operators $s$ by $s_{-}$ and $s_{+}$. The expressions are prevented to become ill-defined by ensuring minimal realizations using a default tolerance of $\sqrt{\epsilon} = 1.5\cdot10^{-8}$.

The obtained models are substituted in Eqs.~\eqref{eq:MBC_AO_TFmatrixD}~and~\eqref{eq:MBC_AO_TFmatrixO}. The system interconnection measure $R_\mathrm{\#}$ is evaluated at $\omega=10^{-2}$~rad\,s\textsuperscript{-1} for each linear model $k$ at a range of azimuth offsets. Because $k$ models are obtained, the optimal offset is defined as the median of computed optimal offsets for each set of linear models. In Figure~\ref{fig:R_BoxPlotAziOff}, the results of the two distinct transformations are presented and compared to the results from spectral analysis in Figure~\ref{fig:MBC_ProblemStatement_OptOffset}. The linear prediction of the optimal azimuth offset including the rotor model cross terms clearly outperforms the case excluding the terms. The provided frequency-domain analysis framework, taking into account blade dynamic coupling, is able to provide a concise estimate of the actual optimal azimuth offset.

\section{Assessment on decoupling and SISO controller design}\label{sec:ASS}
This section investigates the potential application of single-gain and decoupled SISO control loops for IPC by incorporating the optimal azimuth offset. The former aspect is explored using a sensitivity analysis in Section~\ref{sec:S}, whereas the latter aspect is investigated using the Gershgorin circle theorem in Section~\ref{sec:GB}.

\subsection{Sensitivity analysis using singular values plots}\label{sec:S}
In this section the effect of the azimuth offset to the sensitivity function is assessed. The sensitivity function using negative feedback is defined as
\begin{align}
\boldsymbol{S}(j\omega) &= \left( \boldsymbol{I}_{2}+\boldsymbol{L}(j\omega) \right)^{-1},
\end{align}
where $\boldsymbol{L}\in\mathbb{R}^{2\times2}$ is the open-loop gain, which is defined as the multiplication of the multivariable system and the diagonal controller
\begin{align}
\boldsymbol{L}(s) = \tilde{\boldsymbol{P}}(s,\omega_\mathrm{r},\psi_\mathrm{o})\boldsymbol{C}(s),
\end{align}
where $\boldsymbol{C}(s) = \text{diag}\left(c_1(s),~c_2(s)\right)$ consists out of the pure integrators $c_1(s) = c_2(s) = c_\mathrm{I}/s$. For MIMO systems, the sensitivity function gives information on the effectiveness of control through the bounded ratio
\begin{align}
\underset{\bar{}}{\sigma}\left(S(j\omega)\right) \leq \frac{||y(\omega)||_{2}}{||v(\omega)||_{2}} \leq \bar{\sigma}(S(j\omega)),
\end{align}
where $\underset{\bar{}}{\sigma}\left(S(j\omega)\right)$ indicates the smallest and $\bar{\sigma}\left(S(j\omega)\right)$ the highest singular value of $S(j\omega)$, determined by the direction of the output and measurement disturbance signals $y$ and $v$, respectively. For evaluation of the considered MIMO system sensitivity, the singular values of the system frequency response are computed. This is done by performing a Singular Value Decomposition (SVD) on the frequency response of the dynamic system \citep{ref:skogestad2007multivariable}.
\begin{table}[b!]
	\centering
	\caption{The integrator gains $c_\mathrm{I}$ are corrected for the influence of the azimuth offset in the steady-state gain to obtain a consistent control bandwidth.}
	\label{tab:integrator_gains_correction}
	\begin{tabular}{cccccl}
		\hline
		$\boldsymbol{\psi_\mathrm{o}}$  & 0 & 30 & 44\textsuperscript{*} & 58& {deg}\\
		\hline
		${c_\mathrm{I}}$ $\times10^{-6}$ & $3.65$ & $2.66$ & $2.65$ & $2.66$ & rad\,(Nm\,s)\textsuperscript{-1}\\
		\hline
	\end{tabular} 
\end{table}

The sensitivity is evaluated in the fixed frame for the cases without and with the optimal offset. As the offset influences the steady-state gain of the main-diagonal elements, an integral gain correction is applied when implementing an azimuth offset, which is summarized in Table~\ref{tab:integrator_gains_correction}. In this way, a consistent open-loop baseline control bandwidth of $2.2\cdot10^{-2}\times2\pi$~rad\,s\textsuperscript{-1} is attained. It is concluded that the absolute steady-state gain of the main-diagonal terms after transformation with the optimal azimuth offset is increased by $37$~\%.
\begin{figure}[!t]
	\centering
	\includegraphics[scale=1.0]{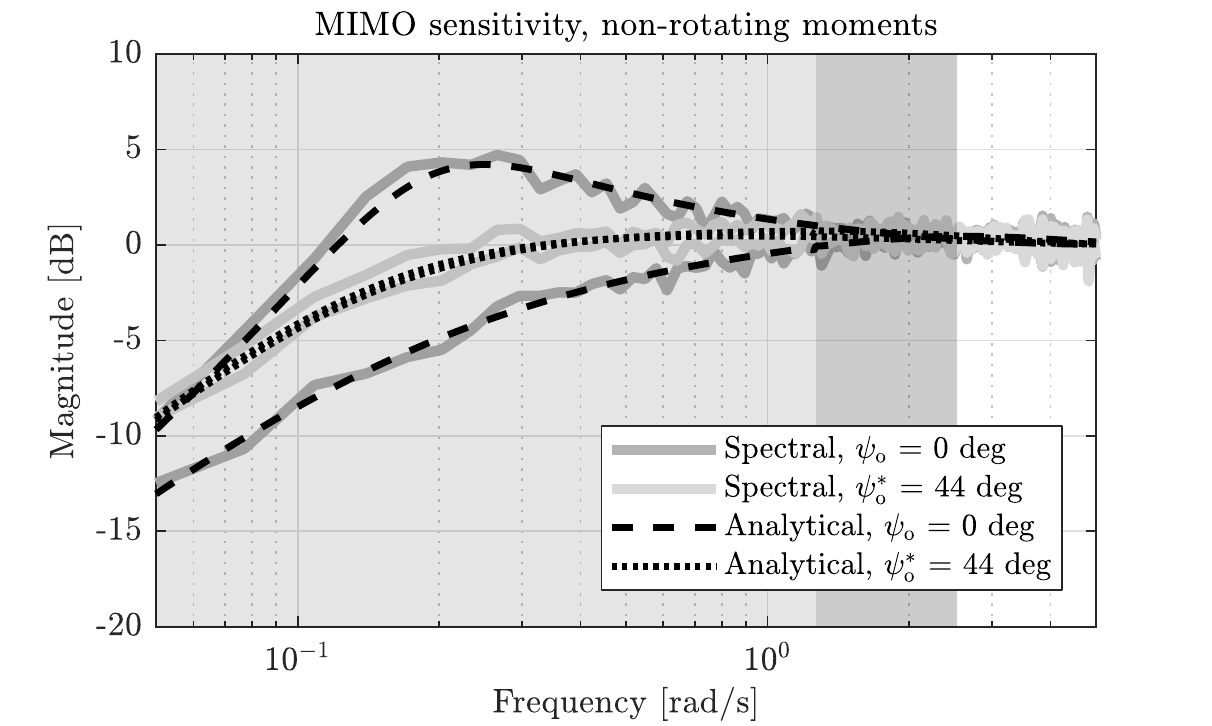}
	\caption{Analysis of azimuth offset on the closed-loop sensitivity in the non-rotating frame, including the diagonal gain-corrected controller $\boldsymbol{C}(s)$. The optimal offset reduces the sensitivity peak and compensates for the gain difference between the trajectories.}%
	\label{fig:S_Sensitivity}%
\end{figure}

Figure~\ref{fig:S_Sensitivity} shows the evaluation of the multivariable sensitivity. The results presented are obtained from high-fidelity simulations (spectral estimate) and from analytical results using the framework presented in this paper. The trajectories show good resemblance for both cases. For the case without an azimuth offset, the peak of the sensitivity function $M_\mathrm{s} = \max_{0\leq\omega<\infty}\left|S(j\omega)\right|$ is the highest and a significant gain difference between the minimum and maximum sensitivity trajectory is observed. On the contrary, the optimal offset results in a smoothened trajectory and an attenuated sensitivity peak, resulting in a more robust IPC implementation. Furthermore, the minimized gain difference reduces directionality and advocates the applicability of decoupled SISO control loops. The gray-shaded regions $\left\{0,~\omega_\mathrm{r}\right\}$ and $\left\{\omega_\mathrm{r},~2\omega_\mathrm{r}\right\}$ are used in Section~\ref{sec:SIMU} for comparison to the rotating blade moments.

\subsection{Decoupling and stability analysis using Gershgorin bands}\label{sec:GB}
Up to this point, a quantification and visualization of the system's degree of decoupling has only been given on simplified linear models using the RGA. For a decoupling and stability analysis of the obtained higher order linearizations, in this section, the Gershgorin circle theorem is employed. The theorem provides both qualitative and quantitative measures of the beforementioned criteria by graphical interpretations and scalar stability margins.

The Gershgorin circle theorem makes use of the Nyquist array containing Nyquist curves of its frequency dependent elements \citep{ref:Maciejowski1989MVFeedbackDesign}. Here, the Nyquist array $\boldsymbol{L}(s)\in\mathbb{R}^{m\times m}$ consists of open loop-transfer elements $l_{ij}(s)$ with $\left\{i,~j\right\}\subset\mathbb{Z}^{m}=\left\{1,~2\right\}$. Furthermore, a Gershgorin band consists of frequency dependent Gershgorin circles with a radius $\mathcal{R}_i(j\omega)$ drawn on the diagonal Nyquist curves $l_{ii}(j\omega)$, defined by
\begin{align}\label{eq:gershgorinradius}
\mathcal{R}_i(j\omega) &= \sum^m_{i,~i\neq j}\left|{l_{ij}(j\omega)}\right|.
\end{align}
Put differently, these bands show the cumulative gains of the row-wise off-diagonal elements of $\boldsymbol{L}(s)$ projected on the main-diagonal Nyquist curves. In general, the off-diagonal Nyquist curves are disregarded for convenient presentation. The closed-loop stability is determined by the Direct Nyquist Array (DNA) stability theorem \citep{ref:Rosenborck1970SSandMVTheory, ref:rosenbrock1976computer}. If the Gershgorin bands do not include the critical $-1$ point, the system is said to be diagonally dominant. The smaller the bands, the higher the diagonal dominance degree, and the system may be treated as $m$ individual SISO systems with negligible interactions. For this reason, the Gershgorin bands can be used as a measure of MIMO (de)coupling \citep{ref:Maciejowski1989MVFeedbackDesign}.

\begin{figure}[t!]
	\centering
	\includegraphics[scale=0.38]{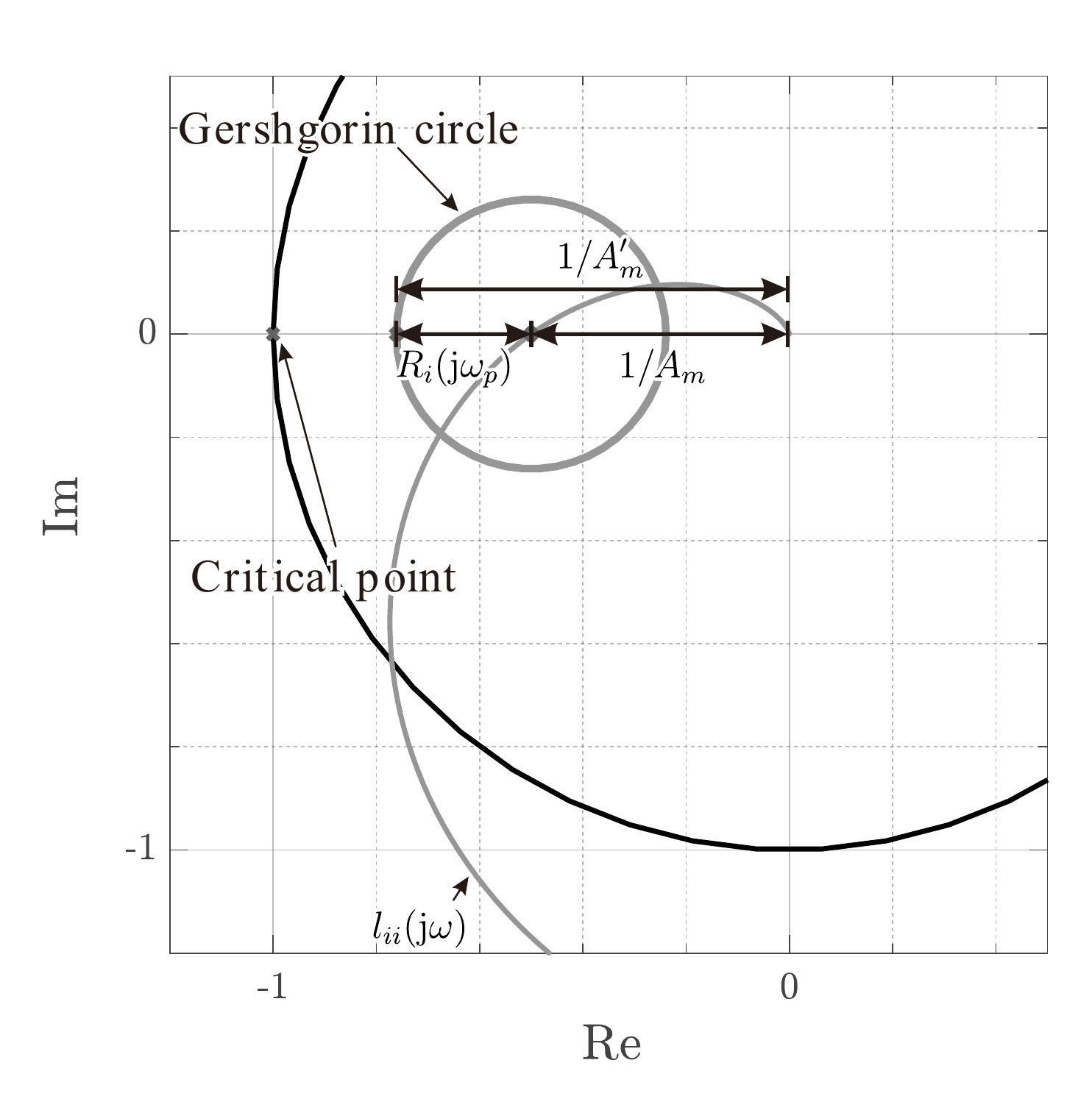}
	\includegraphics[scale=0.38]{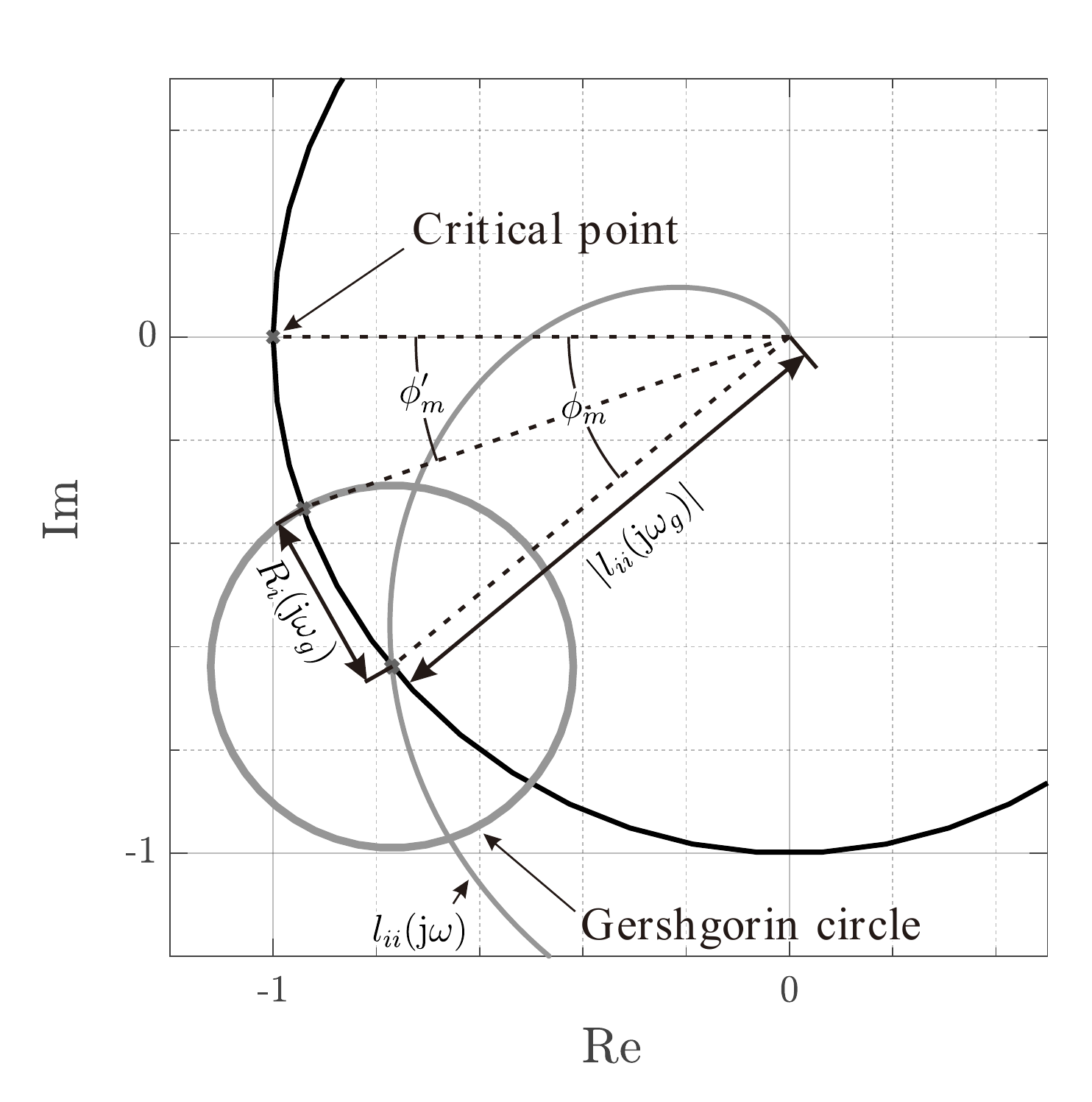}
	\includegraphics[scale=0.38]{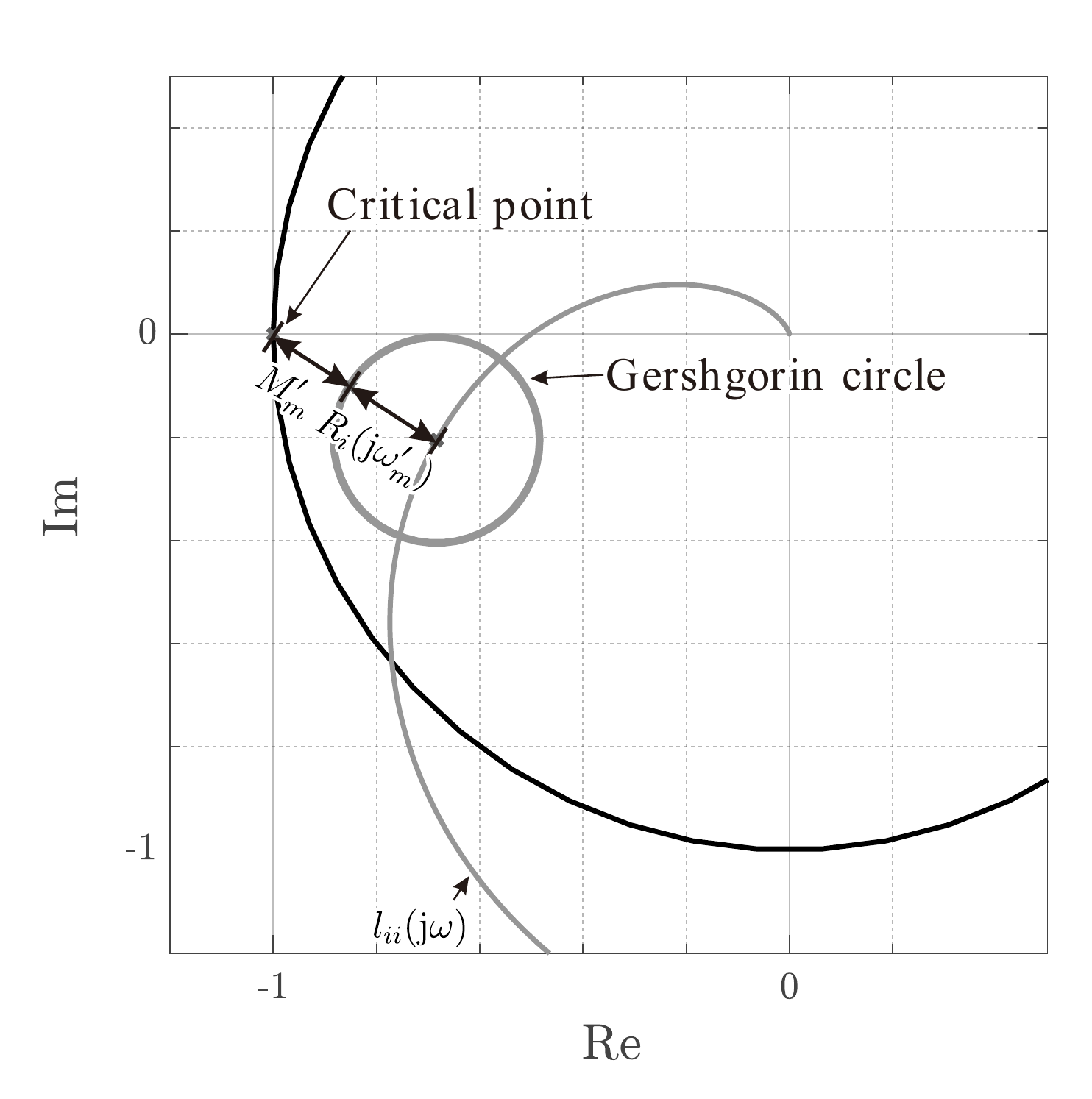}
	\caption{Graphical interpretations of the extended gain margin $A_\mathrm{m}'$ (left), phase margin $\phi_\mathrm{m}'$ (middle) and modulus margin $M_\mathrm{m}'$ (right), adapted from \citep{ref:ho1997tuning, ref:garcia2005pid}. The presence of the Gershgorin circles over the Nyquist locus alters the definition of the conventional margins.}
	\label{fig:Gershgorin_GMPM}%
\end{figure}

Furthermore, Gershgorin bands can be used to shape the earlier defined loop-transfer matrix $\boldsymbol{L}(s)$ according to gain, phase and modulus margins specifications established for SISO controller design. However, due to the presence of the Gershgorin bands over the Nyquist loci, the introduced margins need to be redefined into their \textit{extended} forms \citep{ref:ho1997tuning,ref:garcia2005pid}, denoted by $(\cdot){'}$. Figure~\ref{fig:Gershgorin_GMPM} visualizes the presented notions, and the adapted definitions for gain margin $A_\mathrm{m}$, phase margin $\phi_\mathrm{m}$ and modulus margin $M_\mathrm{m}$ are defined as
\begin{align}
A_\mathrm{m}' &= \frac{A_\mathrm{m}}{\left( 1 + \frac{\sum^m_{i=1,i\neq j}\left|l_{ji}(j\omega_\mathrm{p})\right|}{\left|l_{ii}(j\omega_\mathrm{p})\right|}\right)},\\
\phi_\mathrm{m}' &= \phi_\mathrm{m} - 2\arcsin{\left( \frac{\sum^m_{i=1,i\neq j}\left|l_{ji}(j\omega_\mathrm{g})\right|}{2\left|l_{ii}(j\omega_\mathrm{g})\right|} \right)},\\
M_\mathrm{m}' &= \left|1+l_{ii}(j\omega_\mathrm{m})\right| - \sum_{i,i\neq j}^{m}\left|l_{ji}(j\omega_\mathrm{m})\right|,
\end{align}
where $\omega_\mathrm{p}$, $\omega_\mathrm{g}$ and $\omega_\mathrm{m}$ indicate the frequencies at which the margins are defined. The modulus margin quantifies the sensitivity of the closed-loop system to variations of the considered loop-gain, and thus serves as a measure for robustness. The modulus margin is in general considered as a combined measure of the gain and phase margins, as it represents the minimal distance of the Nyquist locus to the critical $-1$ point by a single value. Consequently, the modulus margin is taken as the main performance indicator in the next section.
\begin{table}[b!]
	\centering
	\caption{The extended gain, phase, and modulus margins of the system of different $\psi_\mathrm{o}$'s. The margins higher than the benchmark ($\psi_\mathrm{o} = 0^\circ$) are underlined. The tilt and yaw loops are denoted by $l_{11}(s)$ and $l_{22}(s)$, respectively.}
	\label{tab:Gershgorin_l1122}
	\begin{tabular}{lcccccc}
		\hline
		\multirow{2}{*}{$\boldsymbol{\psi_\mathrm{o}}$ \textbf{($\boldsymbol{^\circ}$)}} & \multicolumn{2}{c}{$\boldsymbol{A_\mathrm{m}'}\,(-)$} & \multicolumn{2}{c}{\textbf{$\boldsymbol{\phi_\mathrm{m}'}\,(\boldsymbol{^\circ})$}} & \multicolumn{2}{c}{\textbf{$\boldsymbol{M_\mathrm{m}'}\,(-)$}} \\ \cline{2-7}
		& $\boldsymbol{l_{11}(s)}$ & $\boldsymbol{l_{22}(s)}$ & $\boldsymbol{l_{11}(s)}$ & $\boldsymbol{l_{22}(s)}$ & $\boldsymbol{l_{11}(s)}$ & $\boldsymbol{l_{22}(s)}$ \\
		\hline
		0 & -- & -- & -- & -- & -- & -- \\ 
		30 & \underline{23.540} & \underline{23.540} & 71.339 & 71.339 & 0.897 & 0.897 \\
		44 & 21.167 & 21.167 & \underline{84.195} & \underline{84.195} & \underline{0.912} & \underline{0.912} \\
		58 & 18.194 & 18.194 & 71.215 & 71.215 & 0.883 & 0.883 \\
		\hline
	\end{tabular} 
\end{table}
\begin{figure}[t!]
	\centering
	\includegraphics[scale=1.0]{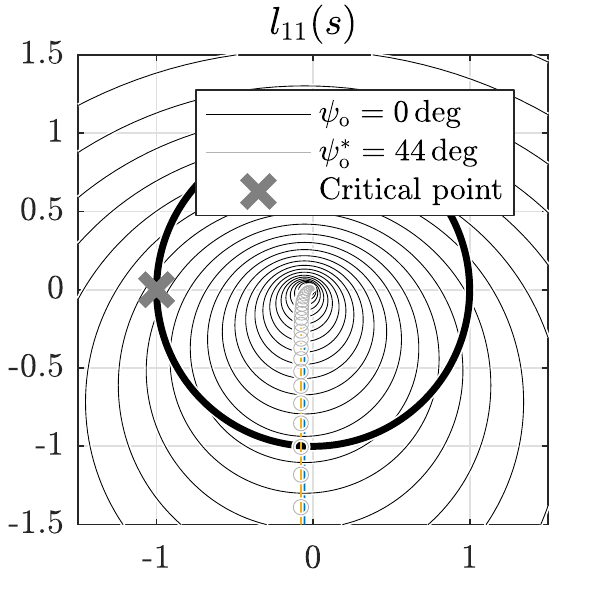}
	\caption{Nyquist loci with Gershorin bands of $l_{11}(s)$. The amount of coupling is greatly reduced and the open-loop system becomes diagonally dominant by incorporating the optimal azimuth offset.}
	\label{fig:Gershgorin_l11}%
\end{figure}

\subsubsection{Decoupling assessment by Gershgorin bands}\label{sec:GB_ASS}
This section assesses and quantifies the degree of decoupling and stability of the IPC implementation for high-order linear models. For this purpose, the Gershgorin circle theorem is used in conjunction with the previously introduced extended margins. The cases considering and disregarding the optimal azimuth offset are examined. 

The first step is to design a compensator that decouples the MIMO system to some extent \citep{ref:ho1997tuning}. For this purpose, the azimuth offset is used, whereafter an actual diagonal controller $\boldsymbol{C}(s)$ is implemented that shapes the loop-gain to attain closed-loop performance and stability specifications. 

Figure~\ref{fig:Gershgorin_l11} shows the Nyquist locus of the first diagonal elements $l_{11}(s)$ using a pure-integrator controller, with and without optimal azimuth offset. The no-offset case has no diagonal dominance, whereas by inclusion of the optimal offset the open-loop system becomes diagonally dominant, shown by the decreased circle radii. In Table~\ref{tab:Gershgorin_l1122} the effect is further quantified by evaluation of the extended stability margins. Two additional (but suboptimal) cases of $30$ and $58$~deg offset are evaluated, and the resulting best margins are underlined. It is shown that the suboptimal case of $30$~deg gives the highest extended gain margins, whereas the optimal offset of $44$~deg results in significantly improved extended phase and modulus margins compared to the baseline case. As the latter mentioned margin is inversely proportional to the sensitivity peak and serves as a main performance indicator, it is concluded that the offset of $44$~deg results in optimal decoupling and robustness.
\begin{figure}[!b]
	\centering
	\includegraphics[scale=1.0]{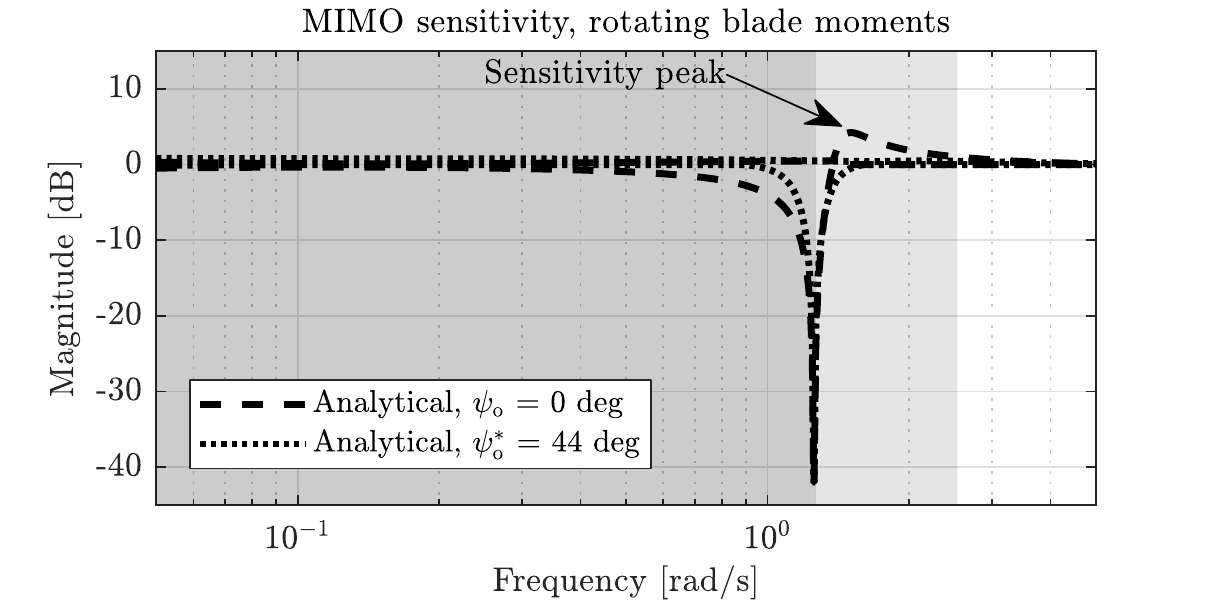}
	\caption{Multivariable sensitivity of the rotating blade moments with and without optimal azimuth offset. The maximum sensitivity peak in the light-gray area is attenuated. The gray-shaded regions relate the sensitivities in the (non-)rotating frames.}
	\label{fig:S_SensitivityRot}%
\end{figure}

\section{Evaluation on the effects of blade load and pitch signals}\label{sec:SIMU}
In this final section, open-loop and closed-loop high-fidelity simulations are performed to evaluate the effect of the azimuth offset on pitch actuation and the blade loads in the rotating frame. The set-up depicted in Figure~\ref{fig:MBC_ClosedLoop} is implemented, and the blade load signal $M_\mathrm{1}$ is recorded. For the closed-loop simulations, a diagonal integral controller $\boldsymbol{C}(s)$ with gains $c_\mathrm{I}$ according to Table~\ref{tab:integrator_gains_correction} is used; for the open-loop simulations the integral gain is set to $0$. A wind profile of $25$~m\,s\textsuperscript{-1} with a Kaimal {IEC~61400$\text{-}$1~Ed.3} turbulence spectrum is used \citep{ref:jonkman2009turbsim}.

\begin{figure}[b!]
	\centering
	\includegraphics[scale=1.0]{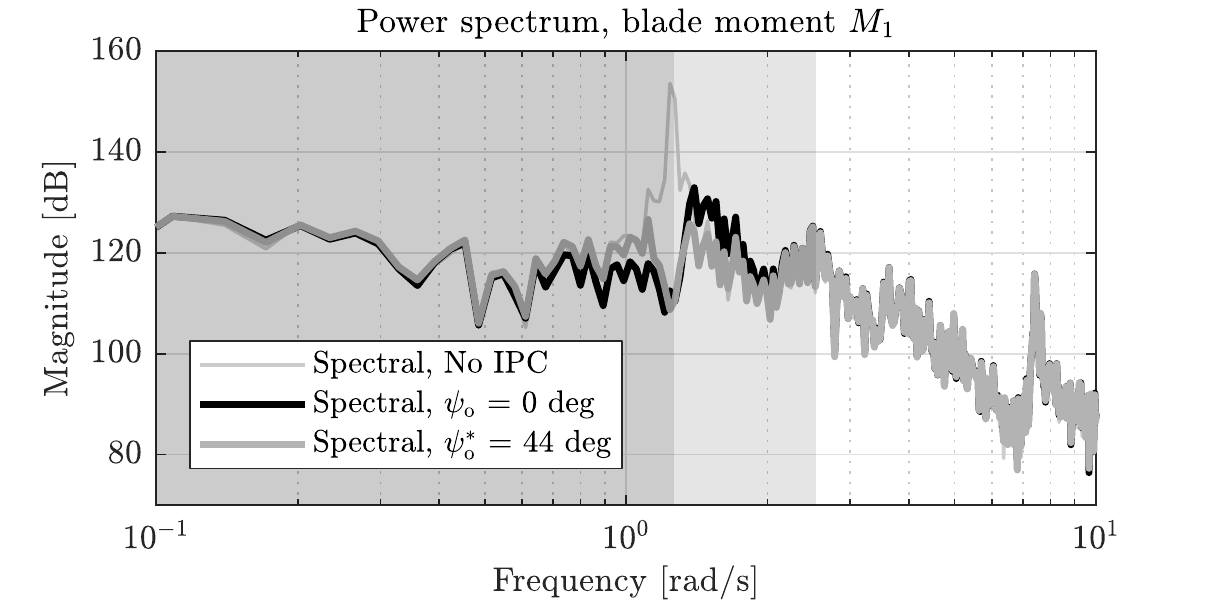}
	\caption{Power spectra of the out-of-plane blade loads, compared for the cases of No IPC, without and with optimal azimuth offset. A significant difference is observed in the light-gray shaded region, where the frequency content significantly drops by inclusion of the offset. For the dark-shaded lower frequency region, the frequency content is slightly increased, however, a more consistent reduction around $1$P is attained.}
	\label{fig:S_RootMOoPCompare}%
\end{figure}
\begin{figure}[!b]
	\centering
	\includegraphics[scale=1.0]{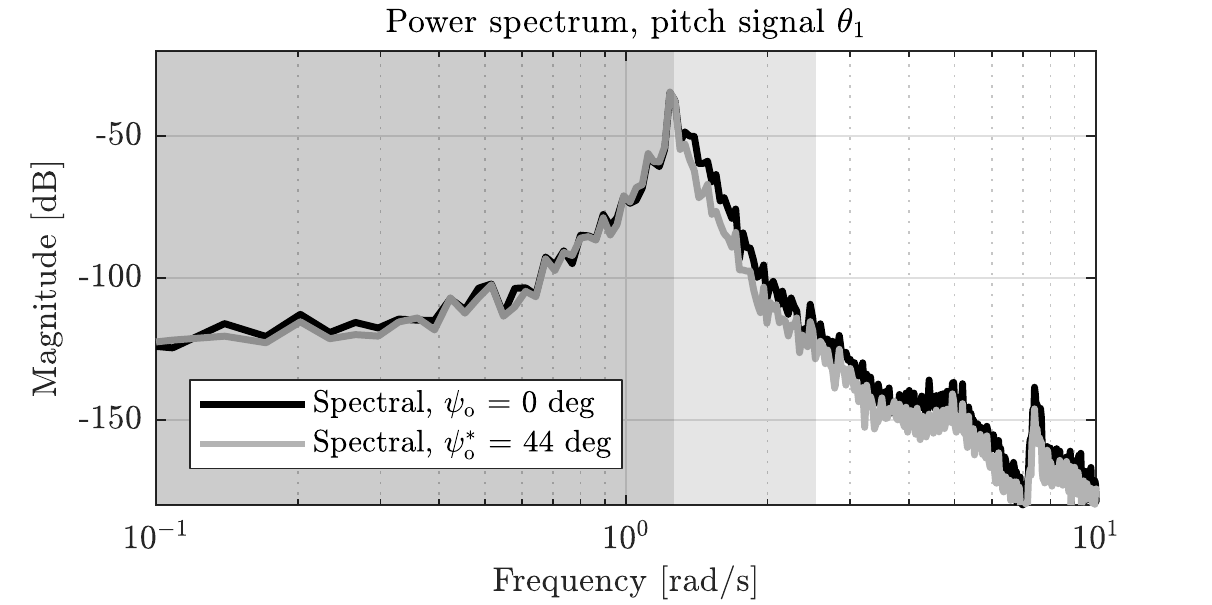}
	\caption{Power spectra of the IPC pitch contribution $\theta_1$, showing a significant overall decrease of high frequency content.}
	\label{fig:S_ThetaMBC1_Compare}%
\end{figure}

Figure~\ref{fig:S_SensitivityRot} presents the multivariable sensitivity of the rotating blade moments for both offset cases. By inclusion of the optimal offset, it is shown that the maximum sensitivity peak around $1.5$~rad\,s\textsuperscript{-1} is attenuated, while the low frequent sensitivity is overall slightly amplified. The same results are observed for the blade moment $M_1$ spectra in Figure~\ref{fig:S_RootMOoPCompare}, resulting in a more consistent reduction of the $1$P load region. By evaluation of the IPC pitch contribution signal $\theta_1$ in Figure~\ref{fig:S_ThetaMBC1_Compare}, it is concluded that the high-frequency actuation content is overall significantly reduced.

Furthermore, the gray-shaded regions of Figure~\ref{fig:S_Sensitivity} and the figures included in this section are interchanged, and indicate the relation between the frequency content in the non-rotating and rotating domains. Referring back to Eq.~\eqref{eq:MBC_FD_revMBC}, the operators $s_\mathrm{+}$ and $s_\mathrm{-}$ show that the frequency content in the rotating domain is mapped from the non-rotating domain by a $1$P shift. Figures~\ref{fig:S_Sensitivity}~and~\ref{fig:S_SensitivityRot} are used for illustration: the peak in the rotating domain at $\omega = 1.5$~rad\,s\textsuperscript{-1} (light-gray) is shifted frequency content from the non-rotating domain at $\omega = 1.5 - 1\mathrm{P} \approx 0.25$~rad\,s\textsuperscript{-1}. 

\section{Conclusions}\label{sec:CONCLUSIONS}  
Although the inclusion of an azimuth offset in the reverse MBC transformation is widely applied in literature, up until now, no profound analysis of its implications has been performed. The analysis in this paper has shown that the application of an azimuth offset further decouples the system in the non-rotating reference frame. The offset for optimal decoupling heavily depends on the changing blade dynamics throughout the entire turbine operating window. The coupling between diagonal and off-diagonal dynamics of the rotor model determines the optimal offset value, and a detailed study is conducted on this aspect. By evaluation of the multivariable system singular values, it is shown that the optimal offset reduces the directionality. Moreover, also the degree of coupling is minimized and the system is made diagonally dominant, as shown using Gershgorin circle theorem. In effect, the application of decoupled and single-gain SISO IPC control loops is justified. Reduction of the sensitivity peak in the non-rotating frame results in attenuation of the maximum sensitivity peak for the rotating blade load sensitivity. As the blade inertia of larger turbine rotors increases significantly for higher power ratings, the inclusion of the azimuth offset in SISO IPC control implementations will be of increased importance. 

\clearpage
\appendix
\section{Including the azimuth offset in a state-space representation}\label{sec:AppA}
The state-space system representation with inclusion of the azimuth offset is presented here. The derivation is based on the work by \citep{ref:bir2008MBC} and the corresponding \texttt{MBC3} code \citep{ref:bir2008MBC3Manual}. The \texttt{MBC3} implementation assumes that the dynamics from individual blade pitch angles to blade root out-of-plane bending moments are described as second-order models. This is in accordance with linear systems obtained from the high-fidelity wind turbine simulation software package FAST \citep{ref:FASTv816}. 
The rotating system is related to the non-rotating system by
\begin{align}
X &= \tilde{\boldsymbol{\mathcal{T}}}^{-1}_{n}X_\mathrm{NR}\label{eq:MBC2B_XNR2X}
\end{align}
and
\begin{gather}
\tilde{\boldsymbol{\mathcal{T}}}_{n}^{-1}(\psi + \psi_\mathrm{o}) = \begin{bmatrix}
\boldsymbol{I}_{F\times F} & 0 \\
0 & \tilde{\boldsymbol{T}}_{n}^{-1}(\psi+\psi_\mathrm{o})
\end{bmatrix},
\end{gather}
where $F$ represents the amount of fixed-frame degrees of freedom and $\tilde{\boldsymbol{\mathcal{T}}}^{-1}(\psi+\psi_\mathrm{o})\in\mathbb{R}^{(F+Bm)\times (F+Bm)}$ is a diagonal matrix, where $m$ is the amount of rotating degrees of freedom.  The forward transformation, transforming the rotating out-of-plane blade moments into their non-rotating counterparts, is defined by $\boldsymbol{T}(\psi)$. Now, combining the results, the following relations transform the periodic matrices to a non-rotating reference frame by applying a state-coordinate change
\begin{align}
A &= \begin{bmatrix}
\boldsymbol{\mathcal{T}}_{n}(\psi) & 0\\
0 & \boldsymbol{\mathcal{T}}_{n}(\psi)
\end{bmatrix} {A}^*(\psi)\left(\begin{bmatrix} {\tilde{\boldsymbol{\mathcal{T}}}_{n}^{-1}}(\psi+\psi_\mathrm{o}) & 0 \\ \omega_\mathrm{r}{\boldsymbol{\mathcal{T}}_{n,2}^{-1}} & {\tilde{\boldsymbol{\mathcal{T}}}_{n}^{-1}}(\psi+\psi_\mathrm{o})\end{bmatrix} - \begin{bmatrix} \omega_\mathrm{r}{\boldsymbol{\mathcal{T}}_{n,2}^{-1}} & 0 \\ \omega_\mathrm{r}^2 {\boldsymbol{\mathcal{T}}_{n,3}^{-1}} + \dot{\omega}_\mathrm{r} {\boldsymbol{\mathcal{T}}_{n,2}^{-1}} & 2\omega_\mathrm{r}{\boldsymbol{\mathcal{T}}_{n,2}^{-1}}
\end{bmatrix}\right),\label{eq:MBC2B_ANR}\\
B &= \begin{bmatrix}
\boldsymbol{\mathcal{T}}_{n}(\psi) & 0\\
0 & \boldsymbol{\mathcal{T}}_{n}(\psi)
\end{bmatrix}B^*(\psi){\boldsymbol{\mathcal{T}}_{n}^{-1}}_{\mathrm{c}}(\psi+\psi_\mathrm{o}),\label{eq:MBC2B_BNR}\\
C &= \boldsymbol{\mathcal{T}}_{n,\mathrm{o}}(\psi)\begin{bmatrix} C_1^*(\psi){\boldsymbol{\mathcal{T}}_{n}^{-1}}(\psi+\psi_\mathrm{o})+\omega_\mathrm{r}C_2^*(\psi){\boldsymbol{\mathcal{T}}_{n}^{-1}} & C_2^*(\psi){\boldsymbol{\mathcal{T}}_{n}^{-1}}(\psi+ \psi_\mathrm{o}) \end{bmatrix},\label{eq:MBC2B_CNR}\\
D &= \boldsymbol{\mathcal{T}}_{n,\mathrm{o}}(\psi)D^*(\psi){\boldsymbol{\mathcal{T}}_{n,\mathrm{c}}^{-1}}(\psi+\psi_\mathrm{o}),\label{eq:MBC2B_DNR}
\end{align}
where $\boldsymbol{\mathcal{T}}_\mathrm{2,3}$ are the first and second time derivative of $\boldsymbol{\mathcal{T}}$, independent of the azimuth offset $\psi_\mathrm{o}$. The $(\cdot)^*$ notation refers to the system $A$, input $B$, output $C$ and feed-through $D$ matrices defined in the rotating frame, and the matrices $A^*\in\mathbb{R}^{r\times r}$ and $C^*\in\mathbb{R}^{q\times r}$ are partitioned as
\begin{gather}
A^*(\psi) = \begin{bmatrix}0 & I\\ A^*_\mathrm{K}(\psi) & A^*_\mathrm{C}(\psi)\end{bmatrix},\\ C^*(\psi) = \begin{bmatrix}C^*_1(\psi) & C^*_2(\psi)\end{bmatrix}.
\end{gather}
As it is assumed that the rotating linearized models only include in- and outputs corresponding to rotating degrees of freedom, the matrices $\boldsymbol{\mathcal{T}}_{\mathrm{c}}^{-1}$ and $\boldsymbol{\mathcal{T}}_{\mathrm{o}}$ are equal to $\boldsymbol{\mathcal{T}}^{-1}$. For obtaining the forward transformation matrix, the inverse matrices $\boldsymbol{\mathcal{T}}^{-1}$, $\boldsymbol{\mathcal{T}}_{\mathrm{c}}^{-1}$ and $\boldsymbol{\mathcal{T}}_{\mathrm{o}}^{-1}$ are required.

\clearpage
\bibliography{REFERENCES_WES2018AziOffset}%

\begin{thebibliography}{10}
\providecommand \doibase [0]{http://dx.doi.org/}%

\bibitem{ref:caselitz1997reduction}
Caselitz P, Kleinkauf W, Kr{\"u}ger T, Petschenka J, Reichardt M, St{\"o}rzel
  K. Reduction of fatigue loads on wind energy converters by advanced control
  methods. {\it EWEC} 1997.

\bibitem{Upwind2006Report}
Fischer T. {Integrated Wind Turbine Design - Final report Task 4.1}. tech.
  rep., Project UpWind;  2006.

\bibitem{park1929two}
Park RH. Two-reaction theory of synchronous machines generalized method of
  analysis-part I. {\it Transactions of the American Institute of Electrical
  Engineers} 1929\string; 48(3)\string: 716--727.

\bibitem{johnson2012helicopter}
Johnson W. {\it Helicopter theory}.
\newblock Courier Corporation .
\newblock 2012.

\bibitem{ref:Menezes2018WTCReview}
Menezes EJN, Ara{\'u}jo AM, {da Silva} NSB. A review on wind turbine control
  and its associated methods. {\it Journal of Cleaner Production} 2018\string;
  174\string: 945--953.

\bibitem{ref:bossanyi2013validation}
Bossanyi EA, Fleming PA, Wright AD. Validation of individual pitch control by
  field tests on two-and three-bladed wind turbines. {\it IEEE Transactions on
  Control Systems Technology} 2013\string; 21(4)\string: 1067--1078.

\bibitem{ref:solingen2016field}
Solingen E, Fleming PA, Scholbrock A, Wingerden J. Field testing of linear
  individual pitch control on the two-bladed controls advanced research
  turbine. {\it Wind Energy} 2016\string; 19(3)\string: 421--436.

\bibitem{ref:shan2013field}
Shan M, Jacobsen J, Adelt S. Field testing and practical aspects of load
  reducing pitch control systems for a {5 MW} offshore wind turbine. {\it
  Annual Conference and Exhibition of European Wind Energy Association} 2013.

\bibitem{ref:jelavic2010estimation}
Jelavi{\'c} M, Petrovi{\'c} V, Peri{\'c} N. Estimation based individual pitch
  control of wind turbine. {\it Automatika} 2010\string; 51(2)\string:
  181--192.

\bibitem{ref:bossanyi2003individual}
Bossanyi E. Individual blade pitch control for load reduction. {\it Wind
  energy} 2003\string; 6(2)\string: 119--128.

\bibitem{ref:geyler2007individual}
Geyler M, Caselitz P. Individual blade pitch control design for load reduction
  on large wind turbines. {\it European Wind Energy Conference (EWEC 2007)}
  2007.

\bibitem{ref:navalkar2014sprc}
Navalkar ST, Van~Wingerden J, Van~Solingen E, Oomen T, Pasterkamp E, Van~Kuik
  G. Subspace predictive repetitive control to mitigate periodic loads on large
  scale wind turbines. {\it Mechatronics} 2014\string; 24(8)\string: 916--925.

\bibitem{ref:spencer2013model}
Spencer MD, Stol KA, Unsworth CP, Cater JE, Norris SE. Model predictive control
  of a wind turbine using short-term wind field predictions. {\it Wind Energy}
  2013\string; 16(3)\string: 417--434.

\bibitem{ref:petrovic2015advanced}
Petrovi{\'c} V, Jelavi{\'c} M, Baoti{\'c} M. Advanced control algorithms for
  reduction of wind turbine structural loads. {\it Renewable Energy}
  2015\string; 76\string: 418--431.

\bibitem{bossanyi2009upwind}
Bossanyi E, Witcher D. Controller for 5MW reference turbine. tech. rep.,
  Upwind;  2009.

\bibitem{houtzager2013wind}
Houtzager I, {van Wingerden} J, Verhaegen M. Wind turbine load reduction by
  rejecting the periodic load disturbances. {\it Wind Energy} 2013\string;
  16(2)\string: 235--256.

\bibitem{mulders2015iterative}
Mulders S. Iterative feedback tuning of feedforward IPC for two-bladed wind
  turbines: A comparison with conventional IPC. Master's thesis. {Delft
  University of Technology}.  2015.

\bibitem{lu2015analysis}
Lu Q, Bowyer R, Jones BL. Analysis and design of Coleman transform-based
  individual pitch controllers for wind-turbine load reduction. {\it Wind
  Energy} 2015\string; 18(8)\string: 1451--1468.

\bibitem{ref:Unguran2019FFIPC}
Ungur\'{a}n R, Boersma S, Petrovi\'{c} V, {van Wingerden} JW, Pao LY, Martin K.
  Feedback-feedforward individual pitch control design with uncertain
  measurements. {\it submitted to: American Control Conference} 2019.

\bibitem{ref:bir2008MBC}
Bir G. Multi-blade coordinate transformation and its application to wind
  turbine analysis. {\it 46th AIAA aerospace sciences meeting and exhibit}
  2008.

\bibitem{ref:Disario2018MScThesis}
Disario G. {On the effects of an azimuth offset in the MBC-transformation used
  by IPC for wind turbine fatigue load reductions}. {\it TU Delft} 2018.

\bibitem{ref:FASTv816}
{NREL - NWTC} . {FAST v8.16}. \url{https://nwtc.nrel.gov/FAST8};  2018.
\newblock [Online; accessed 27-August-2018].

\bibitem{ref:Ljung1999SysID}
Ljung L. {\it {System Identification: Theory for the User}}.
\newblock Prentice Hall .
\newblock 1999.

\bibitem{ref:PBSIDToolbox}
{van Wingerden} JW. {PBSID-Toolbox}. 2018.
\newblock https://github.com/jwvanwingerden/PBSID-Toolbox.

\bibitem{ref:skogestad2007multivariable}
Skogestad S, Postlethwaite I. {\it Multivariable feedback control: analysis and
  design}.
\newblock Wiley New York .
\newblock 2007.

\bibitem{Oppenheim2013SignalsSystems}
Oppenheim A, Willsky A, Nawab S. {\it {Signals and Systems}}.
\newblock Pearson .
\newblock 2013.

\bibitem{Stewart2009Calculus}
Stewart J. {\it {Calculus - Early Transcedentals 6E}}.
\newblock Brooks/Cole .
\newblock 2009.

\bibitem{ref:FASTv8GUI}
Bos R, Zaaijer M, Mulders S, {van Wingerden} J. FASTv8GUI.
  \url{https://github.com/TUDelft-DataDrivenControl/FASTv8GUI};  2018.

\bibitem{ref:Maciejowski1989MVFeedbackDesign}
Maciejowski J. {\it Multivariable Feedback Design}.
\newblock Addison-Wesley .
\newblock 1989.

\bibitem{ref:Rosenborck1970SSandMVTheory}
Rosenbrock H. {\it State-Space and Multivariable Theory}.
\newblock Thomas Nelson \& Sons Ltd .
\newblock 1970.

\bibitem{ref:rosenbrock1976computer}
Rosenbrock HH, Owens D. Computer aided control system design. {\it IEEE
  Transactions on Systems, Man, and Cybernetics} 1976(11)\string: 794--794.

\bibitem{ref:ho1997tuning}
Ho WK, Lee TH, Gan OP. {Tuning of Multiloop Proportional- Integral- Derivative
  Controllers Based on Gain and Phase Margin Specifications}. {\it Industrial
  and engineering chemistry research} 1997.

\bibitem{ref:garcia2005pid}
Garcia D, Karimi A, Longchamp R. PID controller design for multivariable
  systems using Gershgorin bands. {\it IFAC Proceedings Volumes} 2005\string;
  38(1)\string: 183--188.

\bibitem{ref:jonkman2009turbsim}
Jonkman BJ. {\it TurbSim user's guide: Version 1.50}.  2009.

\bibitem{ref:bir2008MBC3Manual}
Bir G. {\it {User's Guide to MBC3}}. NREL;  2008.

\end{thebibliography}

\end{document}